\documentclass[journal ]{new-aiaa}
\usepackage[utf8]{inputenc}
\usepackage{textcomp}

\usepackage{graphicx}
\usepackage{amsmath}
\usepackage[version=4]{mhchem}
\usepackage{siunitx}
\usepackage{longtable,tabularx}
\usepackage{color}
\usepackage{bm}
\usepackage{subcaption}
\usepackage{algorithm}
\usepackage{algpseudocode}

\title{Boundary layer stability and transition analysis using the nonlinear One-Way Navier-Stokes approach}

\author{Michael K. Sleeman\footnote{PhD Candidate, Mechanical and Civil Engineering.} and Tim Colonius\footnote{Frank and Ora Lee Marble Professor of Mechanical Engineering and Medical Engineering.}}
\affil{California Institute of Technology, Pasadena, CA, USA}
\author{Matthew T. Lakebrink\footnote{Boeing Associated Technical Fellow.}}
\affil{Boeing Research \& Technology, Hazelwood, MO, 63042, United States}

\begin{document}

\maketitle

\begin{abstract}
We extend the One-Way Navier Stokes (OWNS) approach to support nonlinear interactions between waves of different frequencies, which will enable nonlinear analysis of instability and transition. In OWNS, the linearized Navier-Stokes equations are parabolized and solved in the frequency domain as a spatial initial-value (marching) problem. OWNS yields a reduced computational cost compared to global solvers, while also conferring numerous advantages over the parabolized stability equations (PSE), despite its higher computational cost relative to PSE, that we seek to extend to nonlinear analysis. We validate the nonlinear OWNS (NOWNS) method by examining nonlinear evolution of two- and three-dimensional disturbances in a low-speed Blasius boundary layer compared to nonlinear PSE (NPSE) and direct numerical simulation (DNS) results from the literature. We demonstrate that NOWNS can be used to simulate flows involve blowing/suction strips, is more robust to numerical noise, and converges for stronger nonlinearities, as compared to NPSE.
\end{abstract}

\section*{Nomenclature}

{\renewcommand\arraystretch{1.0}
\noindent\begin{longtable*}{@{}l @{\quad=\quad} l@{}}
$(x,y,z)$ & streamwise, wall-normal, and spanwise coordinates\\
$\nu$ & specific volume\\
$(u,v,w)$ & velocity (streamwise, wall-normal, and spanwise)\\
$p$ & pressure\\
$\mu$ & dynamic viscosity\\
$a$ & sound speed\\
$\mathrm{Ma}$ & Mach number\\
$c_p$ & specific heat capacity at constant pressure\\
$\gamma$ & heat capacity ratio\\
$k$ & thermal conductivity\\
$\delta_0^*$ & inlet Blasius length scale, defined as $\delta_0^*=\sqrt{\frac{x_0^*\mu_\infty^*\nu_\infty^*}{U_\infty^*}}$\\
$Re$ & Reynolds number in terms of $\delta_0^*$, defined as $Re=\frac{a_\infty^* \delta_0^*}{\nu_\infty^*\mu_\infty^*}$\\
$Pr$ & Prandtl number, defined as $Pr=\frac{\delta_0^*\mu_\infty^*}{k^*}$\\
$\omega$ & temporal frequency\\
$F$ & dimensionless temporal frequency, defined as $F=\frac{\omega^*\nu_\infty^*\mu_\infty^*}{U_\infty^{*2}}$\\
$\beta$ & spanwise wavenumber\\
$b$ & dimensionless spanwise wavenumber, defined as $b=\frac{\beta^*\nu_\infty^*\mu_\infty^*}{U_\infty^*}$\\
$\alpha$ & streamwise wave number (complex-valued, where $\alpha_r$ is the phase and $\alpha_i$ is the growth rate)\\
$\mathrm{Re}_x$ & Reynolds number in terms of $x$, defined as $\mathrm{Re}_x=\frac{U_\infty^* x^*}{\nu_\infty^*\mu_\infty^*}$\\
$M$ & denotes the number of temporal Fourier modes\\
$N$ & denotes the number of spanwise Fourier modes\\
$(m,n)$ & denotes Fourier mode with temporal frequency $m\omega$ and spanwise wavenumber $n\beta$\\
\multicolumn{2}{@{}l}{Subscripts}\\
$\infty$ &  free-stream value\\
$mn$ & a quantity associated with temporal frequency $m\omega$ and spanwise wavenumber $n\beta$\\
$+-0$ & plus, minus, and zero characteristics\\
\multicolumn{2}{@{}l}{Superscripts}\\
$*$ & dimensional quantity\\
$\prime$ & a disturbance to the equilibrium solution\\
$\ddagger$ & a matrix or a vector containing the OWNS auxiliary values\\
\multicolumn{2}{@{}l}{Other symbols}\\
$\bar{(\cdot)}$ & a quantity associated with the equilibrium solution (e.g., Blasius base flow)\\
$\overline{(\cdot)}$ & a complex conjugate value\\
$\hat{(\cdot)}$ & a quantity in Fourier space
\end{longtable*}}

\section{Introduction}\label{sec:intro}

Modern industry tools for transition prediction involve extrapolation from linear amplification through the $e^N$-method~\cite{Smith_1956_eN,VanIngen_1956_eN} or the variable $N$-factor approach~\cite{Crouch_2000_NFactor}. Fundamentally, transition can be studied using Direct Numerical Simulation (DNS)~\cite{Fasel_1990_Subharmonic,Rist_1995_Subharmonic,Sayadi_2013_DNS,Wu_2009_DNS} or Large Eddy Simulation (LES)~\cite{Voke_1995_LES,Huai_1997_LES}, but these approaches are limited by their high computational cost. The nonlinear parabolized stability equations (NPSE) entail a much lower computational cost~\cite{Bertolotti_1992_PSE}, but have several limitations. It is well known that PSE does not remove all upstream-going waves, relying on an {\it ad hoc} regularization that leads to a minimum step size for numerical stability, preventing convergence at all streamwise length scales~\cite{Li_1996_PSE,Li_1997_PSE,Broadhurst_2008_PSE}. More vexingly, even in the linear case, it was shown by Towne et al.~\cite{Towne_2019_PSE} that PSE is generally only accurate for disturbances with a single wavelength at each frequency, and that other disturbances are neither necessarily damped away, nor properly evolved. Throughout this paper, we refer to disturbances without a single clear dominant wavelength, for a given frequency, to be \textit{multi-modal} disturbances. Although some authors~\cite{DeTullio_2013_Roughness,Paredes_2016_Streaks,Hack_2017_Transient} report successful computations of \textit{non-modal} disturbances (transient growth), other authors \cite{Cheung_2007_Transient,Rodriguez_2018_Transient} have reported limitations and challenges.

The linear One-Way Navier-Stokes (OWNS) approach overcomes these limitations, but at a greater computational cost. In particular, it properly removes all upstream-going waves so that the regularizations used by PSE are not required, thus allowing it to capture properly non-modal and multi-modal instabilities, and to take arbitrarily small steps to resolve all length scales~\cite{Towne_2019_PSE}. The linear OWNS approach has previously been applied to two-dimensional (2D) and three-dimensional (3D) boundary-layer flows~\cite{Rigas_2017_OWNS_BL}, and has proven useful hypersonic boundary layer flows~\cite{Kamal_2020_HOWNS,Kamal_2021_OWNS_IO,Kamal_2022_HOWNS}. Although PSE can be used to study hypersonic boundary-layer flows, as in Pruett et al.~\cite{Pruett_1995_DNS}, Paredes et al.~\cite{Paredes_2016_Optimal}, or Lakebrink et al.~\cite{Lakebrink_2017_Crossflows}, hypersonic boundary-layer flows have multiple discrete modes and branches of disturbances that can synchronize, which proves challenging for PSE~\cite{Towne_2019_PSE,Fedorov_2011_Hypersonic}.

The nonlinear OWNS (NOWNS) methodology has been validated for Blasius boundary layer flows~\cite{Sleeman_2023_NOWNS-Aviation,Sleeman_2024_NOWNS-Aviation}. In this paper, we refine the NOWNS methodology, validate it, and demonstrate that it is more robust to numerical noise and converges for stronger nonlinearity than NPSE, and that it can be used to replicate DNS studies involving blowing/suction strips.

\section{Method}

The OWNS procedure was first developed as a method for constructing well-posed one-way approximations of linear hyperbolic systems with slowly-varying coefficients in the direction of integration~\cite{Towne_2015_OWNS-O}, allowing the equations to be solved as a spatial-initial-value problem, leading to a reduced computational cost. The framework has also been applied to the Navier-Stokes equations, linearized about a slowly-varying baseflow, to perform linear stability analysis of jets and boundary layer flows~\cite{Towne_2015_OWNS-O,Towne_2022_OWNS-P,Kamal_2020_HOWNS,Kamal_2021_OWNS_IO,Kamal_2022_HOWNS,Rigas_2017_OWNS_BL}. Here we generalize the OWNS methodology to support nonlinear effects.

The non-dimensional, compressible Navier-Stokes equations for an ideal gas can be
written as
\begin{subequations}
\begin{align}
    \frac{D\nu}{Dt}-\nu(\nabla\cdot\bm{u})&=0,\\
    \frac{D\bm{u}}{Dt}+\nu\nabla p&=\frac{1}{Re}\nu\nabla\cdot\tau,\\
    \frac{Dp}{Dt}+\gamma p(\nabla\cdot\bm{u})
    &=\frac{1}{Re}(\gamma-1)[(\nabla\bm{u}):\tau]+\frac{\gamma}{PrRe}(\nu\nabla^2p+2\nabla\nu\cdot\nabla p+p\nabla^2\nu),
\end{align}
\label{eq:nonDimNSE}
\end{subequations}
for the stress tensor
\[
\tau = \mu \Big(\nabla\bm{u}+(\nabla\bm{u})^T\Big) - \Big(\frac{2}{3}\mu-\kappa\Big)(\nabla\cdot\bm{u})I,
\]
where $\nu$ is the specific volume, $\bm{u}$ is the velocity, $p$ is the pressure, $\gamma$ is heat capacity ratio, $\mu$ is the dynamic viscosity and $\kappa$ is the bulk viscosity. We take the bulk viscosity to be zero ($\kappa=0$) and we introduce the Reynolds number, the Prandtl number, and the Blasius length scale
\[
Re=\frac{\delta_0^*a_\infty^*}{\nu_\infty^*\mu_\infty^*},\quad Pr=\frac{c_{p,\infty}^*\mu_\infty^*}{k_{\infty}^*},\quad \delta_0^*=\frac{\nu_\infty^*x_0^*}{U_\infty^*}.
\]
Here $\nu_\infty^*$ denotes the free-stream specific volume, $\mu_\infty^*$ the free-stream dynamic viscosity, $k_\infty^*$ the free-stream thermal diffusivity, $c_{p,\infty}^*$ the free-stream specific heat capacity at constant pressure, $a_\infty^*$ the free-stream sound speed, while $x$, $y$, $z$, correspond to the streamwise, wall-normal and spanwise directions, respectively. At low Mach numbers, temperature fluctuations are minimal, so we take the fluid properties ($\mu$, $k$, and $c_p$) to be constant.

\subsection{Linear OWNS}

The NOWNS procedure follows closely the linear OWNS procedure, which we introduce first. We wish to study the evolution of a time-varying disturbance, $\bm{q}'$, to a time-invariant equilibrium solution, $\bar{\bm{q}}$, so we decompose the flow into terms associated with $\bar{\bm{q}}$ and $\bm{q}'$, as described in appendix~\ref{app:nse}, yielding the stability equations~\eqref{eq:operatorForm}, for the vector of primitive variables $\bm{q}=(\nu,u,v,w,p)$. To simplify the exposition, we temporarily neglect the streamwise diffusion terms by introducing the exogeneous forcing function $\bm{f}=\bm{f}_{\mathrm{forcing}}+\bm{f}_{\mathrm{viscous}}$ where
\[
\bm{f}_{\mathrm{viscous}}=
- B_{x}(\bar{\bm{q}})\frac{\partial\bm{q}'}{\partial x}
- B_{xx}(\bar{\bm{q}})\frac{\partial^2\bm{q}'}{\partial x^2}
- B_{xy}(\bar{\bm{q}})\frac{\partial^2\bm{q}'}{\partial x\partial y}
- B_{xz}(\bar{\bm{q}})\frac{\partial^2\bm{q}'}{\partial x\partial z},
\]
represents streamwise diffusion terms associated with the disturbance variable, while $\bm{f}_{\mathrm{forcing}}$ is an arbitrary forcing function. We then neglect the nonlinear term and re-write equation~\eqref{eq:operatorForm} to obtain the linear stability equation
\begin{equation}
    A_x(\bar{\bm{q}})\frac{\partial\bm{q}'}{\partial x}
    =\mathcal{L}(\bar{\bm{q}})\bm{q}'+\bm{f}.
    \label{02:eq:opFormLin2}
\end{equation}
Previous work OWNS has often neglected $\bm{f}_{\mathrm{viscous}}$ because this simplifies the procedure without significantly impacting its accuracy~\citep{Towne_2015_OWNS-O,Rigas_2017_OWNS_BL,Kamal_2020_HOWNS}, but we have found that this term impacts more significantly the nonlinear calculation, as discussed in appendix~\ref{sec:streamwiseDiffStudy}, so we re-introduce it (approximately) in section~\ref{subsec:linDiff}.

In NOWNS, we consider a system of linear OWNS equations coupled together through the nonlinear term, which acts as an inhomogeneous forcing function taking the place of $\bm{f}$ in~\eqref{02:eq:opFormLin2}. The OWNS outflow (OWNS-O) approach~\cite{Towne_2015_OWNS-O} can only be applied to homogeneous equations, so we must instead consider either the OWNS projection (OWNS-P) approach~\cite{Towne_2022_OWNS-P} or the OWNS recursive (OWNS-R) approach~\cite{Zhu_2021_OWNS-R}. OWNS-R entails a reduced computational cost compared to OWNS-P, but we have found OWNS-P to be more robust, so we use it to develop the NOWNS approach.

\subsubsection{Semi-discrete equations}

Throughout this work, we assume that our disturbances are periodic in the spanwise direction with wavenumber $\beta$ and in time with frequency $\omega$ so that we can expand our disturbances as Fourier series
\begin{equation}
    \bm{q}'(x,y,z,t)=\sum_{m,n=-\infty}^{\infty}\hat{\bm{q}}_{mn}(x,y)e^{i(n\beta z - m\omega t)}.
    \label{eq:qFourierSeries}
\end{equation}
In the linear case, all Fourier modes evolve independently, and we consider a single disturbance of the form $\bm{q}'(x,y,z,t)=\hat{\bm{q}}(x,y)e^{i(n\beta z-m\omega t)}$, while we discretize in the wall-normal direction using a 4th-order central finite differences which we represent using $D\approx\partial/\partial y$. Our semi-discrete linear operator is then
\begin{equation}
    L(\bar{\bm{q}})=
    i\omega I-[A_{y}(\bar{\bm{q}})+B_{y}(\bar{\bm{q}})]D
    -i\beta [A_{z}(\bar{\bm{q}})+B_{z}(\bar{\bm{q}})]-C(\bar{\bm{q}})
    -B_{yy}(\bar{\bm{q}})D^2+\beta^2 B_{zz}(\bar{\bm{q}})-i\beta B_{yz}(\bar{\bm{q}})=0,
    \label{eq:discreteL}
\end{equation}
and we obtain
\begin{equation}
    A_x(\bar{\bm{q}})\frac{\partial\bm{q}'}{\partial x}
    =
    L(\bar{\bm{q}})\bm{q}'+\bm{f},
    \label{eq:discreteEllip}
\end{equation}
a system of ODEs in $x$ comprising $N_v$ variables, where $N_v=5N_y$ ($N_v=4N_y$) in 3D (2D), for the $N_y$ grid points in $y$.

\subsubsection{Parabolization using the OWNS projection procedure}

The above ODEs in $x$ contain components that propagate both upstream and downstream, and cannot be stably integrated without further intervention. Therefore, we remove upstream effects using a projection operator that we apply in the characteristic variables, $\bm{\phi}=T\bm{q}'$, where $T$ are the eigenvectors of $A_x$, while $\tilde{A}_x=TA_xT^{-1}$ are the eigenvalues. Here we have dropped the argument $\bar{\bm{q}}$ for brevity. We transform our equations to characteristic variables as
\begin{equation}
    \tilde{A}_x
    \frac{\partial\hat{\bm{\phi}}}{\partial x}
    =\tilde{L}\hat{\bm{\phi}}
    +\hat{\bm{f}}_{\phi}.
    \label{02:eq:charFormLin}
\end{equation}
with $\tilde{L}=T LT^{-1}-T A_x\frac{\partial T^{-1}}{\partial x}$ and $\hat{\bm{f}}_{\phi}=T\hat{\bm{f}}$, and re-organize the diagonal matrix $\tilde{A}_x$
\begin{equation}
    \tilde{A}_x=\begin{bmatrix}
        \tilde{A}_{++} & 0 & 0\\
        0 & \tilde{A}_{--} & 0\\
        0 & 0 & \tilde{A}_{00}
    \end{bmatrix},
\end{equation}
for the $N_+$ positive eigenvalues $\tilde{A}_{++}>0$, the $N_-$ negative eigenvalues $\tilde{A}_{--}<0$, and the $N_0$ zero eigenvalues $\tilde{A}_{00}=0$, where $N_0+N_-+N_+=N_v$. We further define
\[
    \tilde{A}_{\pm\pm}=
    \begin{bmatrix}
        \tilde{A}_{++} & 0\\
        0 & \tilde{A}_{--}
    \end{bmatrix},\quad
    \tilde{L}_{\pm\pm}
    =
    \begin{bmatrix}
        \tilde{L}_{++} & \tilde{L}_{+-}\\
        \tilde{L}_{-+} & \tilde{L}_{--}
    \end{bmatrix},\quad
    \tilde{L}_{\pm0}
    =
    \begin{bmatrix}
        \tilde{L}_{+0}\\
        \tilde{L}_{-0}
    \end{bmatrix},
    \quad
    \tilde{L}_{0\pm}
    =
    \begin{bmatrix}
        \tilde{L}_{0+}&
        \tilde{L}_{0-}
    \end{bmatrix},
    \quad
    \hat{\bm{\phi}}
    =\begin{bmatrix}
        \hat{\bm{\phi}}_{\pm}\\\hat{\bm{\phi}}_0
    \end{bmatrix},
    \quad
    \hat{\bm{f}}_{\phi}
    =\begin{bmatrix}
        \hat{\bm{f}}_{\phi,\pm} \\ \hat{\bm{f}}_{\phi,0}
    \end{bmatrix},
\]
so that our equations become
\begin{subequations}
    \begin{align}
        \tilde{A}_{\pm\pm}\frac{\partial\hat{\bm{\phi}}_{\pm}}{\partial x}
        &=
        \tilde{L}_{\pm\pm}\hat{\bm{\phi}}_{\pm}+\tilde{L}_{\pm0}\hat{\bm{\phi}}_{0}+\hat{\bm{f}}_{\phi,\pm},\\
        0
        &=
        \tilde{L}_{0\pm}\hat{\bm{\phi}}_{\pm}+\tilde{L}_{00}\hat{\bm{\phi}}_{0}+\hat{\bm{f}}_{\phi,0},
        \label{02:eq:algConstraint}
    \end{align}
    \label{eq:ellipticDAE}
\end{subequations}
which is a differential algebraic equation (DAE) of index 1. We can use the algebraic constraint~\eqref{02:eq:algConstraint} to obtain $\hat{\bm{\phi}}_{0}=-\tilde{L}_{00}^{-1}[\tilde{L}_{0\pm}\hat{\bm{\phi}}_{\pm}+\hat{\bm{f}}_{\phi,0}]$ so that we can re-write our DAE as an ODE
\begin{equation}
    \frac{\partial\hat{\bm{\phi}}_{\pm}}{\partial x}
    =M\hat{\bm{\phi}}_{\pm}+\hat{\bm{g}},
    \label{02:eq:charFormLinODE}
\end{equation}
for $M=\tilde{A}_{\pm\pm}^{-1}[\tilde{L}_{\pm\pm}-\tilde{L}_{\pm0}\tilde{L}_{00}^{-1}\tilde{L}_{0\pm}]$, and $\hat{\bm{g}}=\tilde{A}_{\pm\pm}^{-1}[\hat{\bm{f}}_{\phi,\pm}-\tilde{L}_{\pm0}L_{00}^{-1}\hat{\bm{f}}_{\phi,0}]$.

The upstream- and downstream-going modes of~\eqref{02:eq:charFormLinODE} can be determined based on the eigenvalues of $M$, according to Brigg's criterion~\cite{Briggs_1964_Electron}, which can then be used to introduce well-posed one-way equations, according to the criterion of Kreiss~\cite{Kreiss_1970_IBVP}, as described in Towne and Colonius~\cite{Towne_2015_OWNS-O} and Towne et al.~\cite{Towne_2022_OWNS-P}. We expand the solution  as a linear combination of the eigenvectors
$\hat{\bm{\phi}}_\pm
=V\hat{\bm{\psi}}
=\sum_{k=1}^{N}\bm{v}^{(k)}\hat{\psi}_{k},
$ where $V$ are the eigenvectors $M=VDV^{-1}$.
From Brigg's criterion, we know that $M$ has $N_{+}$ downstream- and $N_{-}$ upstream-going modes, so that we can further partition $V$ into its downstream- ($V_+$) and upstream-going ($V_-$) components yielding the downstream- and upstream-going solutions $\hat{\bm{\phi}}_{\pm}'=V_+\hat{\bm{\psi}}_+$ and $\hat{\bm{\phi}}_{\pm}''=V_-\hat{\bm{\psi}}_-$, respectively. We then use the eigensystem of $M$ to define the projection operator
\[
P=V
\begin{bmatrix}
I_{++} & 0 \\
0 & 0
\end{bmatrix}
V^{-1},
\]
which retains the downstream-going modes associated with $\bm{\psi}_+$ while removing the upstream-going modes associated with $\bm{\psi}_-$. We apply $P$ to our ODE~\eqref{02:eq:charFormLinODE} and use linearity to assert that
\begin{equation*}
    \frac{\partial\hat{\bm{\phi}}'_{\pm}}{\partial x}
    +\frac{\partial\hat{\bm{\phi}}_{\pm}''}{\partial x}
    =P
    [M\hat{\bm{\phi}}_{\pm}+\hat{\bm{g}}]
    +[I-P]
    [M\hat{\bm{\phi}}_{\pm}+\hat{\bm{g}}],
\end{equation*}
yielding the equation for the downstream-going solution
\begin{equation}
    \frac{\partial\hat{\bm{\phi}}_{\pm}'}{\partial x}
    =
    P[M\hat{\bm{\phi}}_\pm'+\hat{\bm{g}}],
    \label{eq:oneWayODE}
\end{equation}
where we used that $P$ and $M$ commute, as shown by Towne et al.~\cite{Towne_2022_OWNS-P}.

\subsubsection{Approximate projection operator}\label{sec:approxProject}

Explicitly using Brigg's criterion to identify upstream- and downstream-going modes in a rigorous way would be computationally expensive and prone to numerical error, so we instead apply the projection operator approximately using a recursive filter~\citep{Towne_2015_OWNS-O,Towne_2022_OWNS-P}.

We define the residuals
$
\hat{\bm{r}}_\pm(\bm{\phi})
=\tilde{A}_{x,\pm\pm}^{-1}[\tilde{L}_{\pm\pm}\hat{\bm{\phi}}_\pm
+\tilde{L}_{\pm0}\hat{\bm{\phi}}_0
+\hat{\bm{f}}_{\phi,\pm}],
$ and
$\hat{\bm{r}}_0(\bm{\phi})
=\tilde{L}_{0\pm}\hat{\bm{\phi}}_\pm
+\tilde{L}_{00}\hat{\bm{\phi}}_0
+\hat{\bm{f}}_{\phi,0}
$
based on \eqref{eq:ellipticDAE}, and following the approach of Towne et al.~\cite{Towne_2022_OWNS-P}, we can apply $P$ approximately to the residual using the recursive filter
\begin{subequations}
\begin{align}
    \hat{\bm{r}}_{+}^{(-N_b)}
    &=0\\
    (\tilde{L}-i\beta_{-}^{(j)}\tilde{A})
    \hat{\bm{r}}^{(-j)}
    -(\tilde{L}-i\beta_{+}^{(j)}\tilde{A})
    \hat{\bm{r}}^{(-j-1)}
    &=0,\quad j=1,\dots,N_b-1\\
    (\tilde{L}-i\beta_{-}^{(0)}\tilde{A})
    \hat{\bm{r}}^{(0)}
    -(\tilde{L}-i\beta_{+}^{(0)}\tilde{A})
    \hat{\bm{r}}^{(-1)}
    &=
    (\tilde{L}-i\beta_{-}^{(0)}\tilde{A})
    \hat{\bm{r}}\\
    \hat{\bm{r}}_0^{(0)}&=\hat{\bm{r}}_0,\label{02:eq:filterConstraint}\\
    (\tilde{L}-i\beta_{+}^{(j)}\tilde{A})
    \hat{\bm{r}}^{(j)}
    -(\tilde{L}-i\beta_{-}^{(j)}\tilde{A})
    \hat{\bm{r}}^{(j+1)}
    &=0,\quad j=0,\dots,N_b-1\\
    \hat{\bm{r}}_{-}^{(N_b)}
    &=0,
\end{align}
\label{eq:approx_project}
\label{02:eq:filterR}
\end{subequations}
where $\hat{\bm{r}}_0=0$ when the algebraic constraint~\eqref{02:eq:algConstraint} is satisfied. 
Here, $\{\beta_{\pm}^{(j)}\}_{j=0}^{N_b-1}$ are termed the \textit{recursion parameters}, while $\{\hat{\bm{r}}^{(j)}\}_{j=-N_b}^{N_b}$ are termed the \textit{auxiliary variables}, where $N_{\mathrm{aux}}=(2N_v+1)N_b$. We introduce the vector auxiliary variables $\hat{\bm{r}}_{\mathrm{aux}}\in\mathcal{C}^{N_{\mathrm{aux}}}$, and the approximate projection operators
$P_1\in\mathbb{C}^{N_{\mathrm{aux}}\times N_v}$, $P_2\in\mathbb{C}^{N_{\mathrm{aux}}\times N_{\mathrm{aux}}}$, $P_3\in\mathbb{R}^{N_{\pm}\times N_{\mathrm{aux}}}$, where $P_1\hat{\bm{r}}$ and $P_2\hat{\bm{r}}_{\mathrm{aux}}$ give the right- and left-hand sides of~\eqref{eq:approx_project}, respectively, while $P_3$ extracts $\hat{\bm{r}}_{\pm}^{(0)}$ from $\hat{\bm{r}}_{\mathrm{aux}}$ as $\hat{\bm{r}}_\pm^{(0)}=P_3\hat{\bm{r}}_{\mathrm{aux}}$. The action of the approximate projection operator on the DAE~\eqref{eq:ellipticDAE} can expressed compactly as
\begin{subequations}
\begin{align}
    \frac{\partial\hat{\bm{\phi}}_\pm'}{\partial x}
    &=P_3\hat{\bm{r}}_{\mathrm{aux}},\\
    P_2\hat{\bm{r}}_{\mathrm{aux}}
    &=P_1
    \begin{bmatrix}
        \tilde{A}_{\pm\pm}^{-1}[
        \tilde{L}_{\pm\pm}\hat{\bm{\phi}}'_\pm
        +\tilde{L}_{\pm0}\hat{\bm{\phi}}'_0
        +\hat{\bm{f}}_{\phi,\pm}
        ]\\
        \tilde{L}_{0\pm}\hat{\bm{\phi}}'_\pm
        +\tilde{L}_{00}\hat{\bm{\phi}}'_0
        +\hat{\bm{f}}_{\phi,0}
    \end{bmatrix}
    ,\\
    0&
    =\tilde{L}_{0\pm}\hat{\bm{\phi}}'_\pm
    +\tilde{L}_{00}\hat{\bm{\phi}}'_0
    +\hat{\bm{f}}_{\phi,0}.
\end{align}
\end{subequations}

\subsubsection{Fully-discrete equations}

We define $A^{\ddagger},L^{\ddagger}\in\mathbb{R}^{(N_v+N_{\mathrm{aux}})\times(N_v+N_{\mathrm{aux}})}$
and $\hat{\bm{\phi}}^{\ddagger},\hat{\bm{f}}_{\phi}^{\ddagger}\in\mathbb{C}^{N_v+N_{\mathrm{aux}}}$ such that
\[
    A^\ddagger
    =
    \begin{bmatrix}
        I_{\pm\pm} & 0 & 0\\
        0 & 0 & 0\\
        0 & 0 & 0
    \end{bmatrix},
    \quad
    L^\ddagger
    =
    \begin{bmatrix}
        0 & 0 & \Delta x P_3\\
        P_1
        \begin{bmatrix}
        \tilde{A}_{\pm\pm}^{-1}L_{\pm\pm}\\
        L_{0\pm}
        \end{bmatrix}
        & P_1
        \begin{bmatrix}
        \tilde{A}_{\pm\pm}^{-1}L_{\pm0}\\
        L_{00}
        \end{bmatrix}
        & -P_2\\
        L_{0\pm} & L_{00}  & 0
    \end{bmatrix},
    \quad
    \hat{\bm{\phi}}^{\ddagger}=
    \begin{bmatrix}
        \bm{\phi}_{\pm}'\\
        \bm{\phi}_{0}\\
        \bm{r}_{\mathrm{aux}}
    \end{bmatrix},
    \quad
    \hat{\bm{f}}_{\phi}^{\ddagger}=
    \begin{bmatrix}
        0\\
        P_1
        \begin{bmatrix}
        \tilde{A}_{\pm\pm}^{-1}\hat{\bm{f}}_{\phi,\pm}\\
        \hat{\bm{f}}_{\phi,0}
        \end{bmatrix}
        \\
        \hat{\bm{f}}_{\phi,0}\\
    \end{bmatrix},
\]
and apply an $s$-order BDF scheme to obtain the linear OWNS equation
\begin{equation}
    [c^{(0)}A^{\ddagger}
    -L^{\ddagger(k+1)}]\hat{\bm{\phi}}^{\ddagger(k+1)} 
    =
    -\sum_{l=1}^{s-1} c^{(l)}A^{\ddagger}\hat{\bm{\phi}}^{\ddagger(k+1-l)}
    +\hat{\bm{f}}_{\phi}^{\ddagger(k+1)},
    \label{02:eq:linSysFD}
\end{equation}
as has been done in previous work on linear OWNS~\cite{Rigas_2017_OWNS_BL,Kamal_2020_HOWNS}.

\subsubsection{Streamwise diffusion terms}\label{subsec:linDiff}

Following discretization in the wall-normal direction and trasnformation to characteristic variables, our streamwise diffusion terms become
\[
\hat{\bm{f}}_{\phi,\mathrm{viscous}}=
-\tilde{B}_{xx}\frac{\partial^2\hat{\bm{\phi}}}{\partial x^2}
-\tilde{B}_{x}\frac{\partial\hat{\bm{\phi}}}{\partial x}
-\tilde{B}\hat{\bm{\phi}}.
\]
for
$\tilde{B}_{xx}=TB_{xx}T^{-1}$,
$\tilde{B}_{x}=2TB_{xx}\frac{\partial T^{-1}}{\partial x}+T[B_{x}+B_{xy}D+i\beta B_{xz}]T^{-1}$, and
$\tilde{B}=TB_{xx}\frac{\partial^2 T^{-1}}{\partial x^2}+T[B_{x}+B_{xy}D+i\beta B_{xz}]\frac{\partial T^{-1}}{\partial x}$. We discretize the second-derivative using a second-order backward difference, while we discretize the first-derivative using the BDF scheme to obtain
\[
\hat{\bm{f}}^{(k+1)}_{\phi,\mathrm{viscous}}=
-\tilde{B}_{xx}\frac{
    \hat{\bm{\phi}}^{(k+1)}
    -2\hat{\bm{\phi}}^{(k)}
    +\hat{\bm{\phi}}^{(k-1)}
    }{(\Delta x)^2}
-\Big(\sum_{l=0}^{s-1} c^{(l)}\tilde{B}_{x}\hat{\bm{\phi}}^{(k+1-l)}\Big)
-\tilde{B}\hat{\bm{\phi}}^{(k+1)},
\]
and we add this term back into the fully-discrete OWNS equations~\eqref{02:eq:linSysFD}.

\subsection{Nonlinear OWNS}~\label{app:detailsNonlinear}

Whereas infinitesimal disturbances (in the linear case) evolve independently from each other so that each Fourier mode can be considered separately, finite amplitude disturbances (in the nonlinear case) are coupled through the nonlinear term. Since it is not feasible to consider an infinite number of Fourier modes, we truncate the Fourier~\eqref{eq:qFourierSeries} series
\begin{equation}
    \bm{q}'(x,y,z,t)=\sum_{m=-M}^{M}\sum_{n=-N}^{N}\hat{\bm{q}}_{mn}e^{i(n\beta z - m\omega t)},
    \label{eq:qFourierSeriesTruncated}
\end{equation}
resulting in $(2M+1)\times(2N+1)$ Fourier modes. However, we require that $\bm{q}'$ be real-valued, which yields the constraint
$\hat{q}_{mn}=\overline{\hat{q}_{-mn}}$,
so that we need only track $(M+1)\times(2N+1)$ Fourier modes. We can additionally introduce a spanwise symmetry condition, as described in section~\ref{sec:sym}, which further reduces the number of modes to $(M+1)\times(N+1)$.

We use our assumption of periodicity to obtain the Fourier series
\[
\tilde{L}(\bar{\bm{q}})\bm{\phi}
=\sum_{m=-M}^{M}\sum_{n=-N}^{N}\hat{L}_{mn}\hat{\bm{\phi}}_{mn}
e^{i(n\beta z-m\omega t)},
\quad
\tilde{\bm{F}}(\bm{\phi})
=
\sum_{m=-M}^{M}
\sum_{n=-N}^{N}
\hat{\bm{F}}_{mn}(\bm{\phi})e^{i(n\beta z-m\omega t)},
\]
where the transformation of $\bm{F}(\bm{\phi})$ to characteristic variables, $\tilde{\bm{F}}(\bm{\phi})$, mimics the transformation of $L(\bar{\bm{q}})$. The Fourier modes are mutually orthogonal, yielding the following system of equations
\begin{equation}
    \tilde{A}\frac{\partial\hat{\bm{\phi}}_{mn}}{\partial x}
    =
    \hat{L}_{mn}\hat{\bm{\phi}}_{mn}
    +\hat{\bm{F}}_{mn}(\bm{\phi})
    +\hat{\bm{f}}_{\phi,mn},
    \quad\forall m\in\mathbb{Z}_M,\quad\forall n\in\mathbb{Z}_N,
    \label{02:eq:nonLinFourier}
\end{equation}
for $\mathbb{Z}_M\equiv \{x\in\mathbb{Z} | -M\leq x\leq M\}$ and $\mathbb{Z}_N\equiv \{x\in\mathbb{Z} | -N\leq x\leq N\}$, where $\mathbb{Z}$ is the set of all integers. We follow a procedure that mimics the linear OWNS approach to obtain
\begin{equation}
    \frac{\partial\bm{\phi}_{\pm,mn}'}{\partial x}
    =
    \hat{P}_{mn}
    [\hat{M}_{mn}\hat{\bm{\phi}}'_{\pm,mn}
    +\hat{\bm{g}}_{mn}(P\bm{\phi}_{\pm}')],
    \quad\forall m\in\mathbb{Z}_M,\quad\forall n\in\mathbb{Z}_N.
    \label{02:eq:nowns}
\end{equation}
In the linear case, $P$ and $M$ commute so that two one-way parabolic equations can be solved to recover the full elliptic solution~\cite{Towne_2022_OWNS-P}. However, this property does not apply in the nonlinear case because the $P$ does not commute with the nonlinear term ($P\bm{g}(\bm{\phi}_{\pm})\neq P\bm{g}(P\bm{\phi}_{\pm})$ in general), so that~\eqref{02:eq:nowns} removes the upstream effect, $\bm{\phi}''$, from the nonlinear term, and when we sum the upstream- and downstream-going equations together, we do not recover the elliptic equation. This is a reasonable choice for convective instabilities, where the disturbances travel primarily in one direction, and we verify $\textit{a posteriori}$ that we match closely DNS results in the literature.

\subsubsection{Fully-discrete equations}

We define $L_{mn}^{\ddagger}\in\mathbb{R}^{(N_v+N_{\mathrm{aux}})\times(N_v+N_{\mathrm{aux}})}$ and $\hat{\bm{\phi}}_{mn}^{\ddagger},\hat{\bm{f}}_{mn,\phi}^{\ddagger}\in\mathbb{C}^{N_v+N_{\mathrm{aux}}}$, as in the linear case, for all $m\in\mathbb{Z}_M$ and $n\in\mathbb{Z}_N$, and further introduce the nonlinear term $\hat{\bm{F}}_{mn}^{\ddagger}\in\mathbb{C}^{N_v+N_{\mathrm{aux}}}$ such that it mimics the definition of the forcing function $\hat{\bm{f}}_{mn,\phi}^{\ddagger}$. Then we discretize the first-derivatives in both the linear and nonlinear terms using the BDF scheme yielding the fully-discrete nonlinear system of equations
\begin{equation}
    \sum_{l=0}^{s-1} c^{(l)}A^{\ddagger}\hat{\bm{\phi}}_{mn}^{\ddagger(k+1-l)}
    =
    \hat{L}_{mn}^{\ddagger(k+1)}\hat{\bm{\phi}}_{mn}^{\ddagger(k+1)}
    +\hat{\bm{F}}_{mn}^{\ddagger(k+1)}
    +\hat{\bm{f}}_{\phi,mn}^{\ddagger(k+1)},\quad
    \forall m\in\mathbb{Z}_M,\quad\forall n\in\mathbb{Z}_N.
    \label{02:eq:nlinSysFD}
\end{equation}

\subsubsection{Pseudo-spectral method}

We employ a pseudo-spectral method whereby we solve our equations in Fourier space, while we evaluate the nonlinear terms in physical space. Given the Fourier coefficients $\hat{\bm{\phi}}_{mn}$ for $m=0,\dots,2M$ and $n=0,\dots,N$, we can use the inverse discrete Fourier transform (iDFT) to compute the solution in physical space as
\begin{equation}
        \bm{\phi}_{mn}
        =
        \frac{1}{2M+1}\frac{1}{2N+1}\sum_{k=0}^{2M}\sum_{l=0}^{2N}
        \hat{\bm{\phi}}_{kl}e^{i2\pi mk/(2M+1)}e^{i2\pi nl/(2N+1)}.
\end{equation}
Then, we can evaluate the nonlinear terms in physical space, $\tilde{\bm{F}}_{mn}=\tilde{\bm{F}}(\bm{\phi}_{mn})$, and employ the discrete Fourier transform (DFT) to compute the Fourier components of the nonlinear terms as
\begin{equation}
    \hat{\bm{F}}_{kl}
    =
    \sum_{m=0}^{2M}\sum_{n=0}^{2N}
    \tilde{\bm{F}}_{mn}e^{-i2\pi mk/(2M+1)}e^{-i2\pi nl/(2N+1)}.
\end{equation}
In practice, the DFT and iDFT are performed using fast Fourier transform (FFT) libraries. At each step, we require the Euclidean norm of the residual for~\eqref{02:eq:nlinSysFD} be converged in both an absolute and relative sense to a tolerance of $10^{-10}$.

\subsubsection{Nonlinear solution procedure and computational cost}\label{subsubsec:compCost}

We can take the derivative of the NOWNS equations with respect to the disturbance variable to obtain the Newton iteration (in 2D with $M=1$)
\begin{equation}
\begin{bmatrix}
        (c^{(0)}A^{\ddagger}-\hat{L}_{0}^\ddagger)
        + \hat{J}^{\ddagger}_{0} &
        \overline{\hat{J}^{\ddagger}_{1}} &
        \hat{J}^{\ddagger}_{1} \\
        \hat{J}^{\ddagger}_{1} &
        (c^{(0)}A^{\ddagger}
        -\hat{L}^{\ddagger}_{1})
        + \hat{J}^{\ddagger}_{0} &
        \overline{\hat{J}^{\ddagger}_{1}} \\
        \overline{\hat{J}^{\ddagger}_{1}} &
        \hat{J}^{\ddagger}_{1} &
        (c^{(0)}A^{\ddagger}
        -\hat{L}_{-1}^\ddagger) + \hat{J}^{\ddagger}_{0}
    \end{bmatrix}
    \begin{bmatrix}
        \Delta\hat{\bm{\phi}}_0^\ddagger \\
        \Delta\hat{\bm{\phi}}_1^\ddagger \\
        \overline{\Delta\hat{\bm{\phi}}_1}^\ddagger
    \end{bmatrix}
    =
    \begin{bmatrix}
        \hat{r}_0^\ddagger \\
        \hat{r}_1^\ddagger \\
        \overline{\hat{r}_1}^\ddagger
    \end{bmatrix},
    \tag{\ref{eq:newton}}
\end{equation}
as discussed in appendix~\ref{app:nownsJac}, where $J_{p-m}^{\ddagger}=-\partial\hat{\bm{F}}^{\ddagger}_m/\partial\hat{\bm{\phi}}_p^{\ddagger}$. If we neglect $J_m^{\ddagger}$ for $m\neq0$, then we obtain
\begin{equation}
    \begin{bmatrix}
        (c^{(0)}A^{\ddagger}-\hat{L}_{0}^\ddagger)
        + \hat{J}_{0}^{\ddagger} &
        0 &
        0 \\
        0 &
        (c^{(0)}A^{\ddagger}-\hat{L}_{1}^\ddagger)
        + \hat{J}_{0}^{\ddagger} &
        0 \\
        0 &
        0 &
        (c^{(0)}A^{\ddagger}-\hat{L}_{-1}^\ddagger)
        + \hat{J}_{0}^{\ddagger}
    \end{bmatrix}
    \begin{bmatrix}
        \Delta\hat{\bm{\phi}}_0^\ddagger \\
        \Delta\hat{\bm{\phi}}_1^\ddagger \\
        \overline{\Delta\hat{\bm{\phi}}_1}^\ddagger
    \end{bmatrix}
    =
    \begin{bmatrix}
        \hat{r}_0^\ddagger \\
        \hat{r}_1^\ddagger \\
        \overline{\hat{r}_1}^\ddagger
    \end{bmatrix}.
\end{equation}
An extension to 3D for Newton's method~\eqref{eq:newton} is straightforward, while system~\eqref{eq:quasiNewton} can be written in 3D as
\begin{equation}
    [c^{(0)}A^{\ddagger}-\hat{L}_{mn}^\ddagger+\hat{J}_{0}^{\ddagger}]
    \hat{\bm{\phi}}_{mn}^{\ddagger(k+1)}
    =
    -\sum_{l=1}^{s-1} c^{(l)}A^{\ddagger}\hat{\bm{\phi}}_{mn}^{\ddagger(k+1-l)}
    +\hat{\bm{F}}_{mn}^{\ddagger(k+1)}
    +\hat{\bm{f}}_{mn}^{\ddagger(k+1)},
    \label{eq:quasiNewton}
\end{equation}
for $m=-M,\dots,M$ and $n=-N,\dots,N$. If we neglect $J^{\ddagger}_m$ entirely, then we obtain
\begin{equation}
    [c^{(0)}A^{\ddagger}-\hat{L}_{mn}^\ddagger]\hat{\bm{\phi}}_{mn}^{\ddagger(k+1)}
    =
    -\sum_{l=1}^{s-1} c^{(l)}A^{\ddagger}\hat{\bm{\phi}}_{mn}^{\ddagger(k+1-l)}
    +\hat{\bm{F}}_{mn}^{\ddagger(k+1)}
    +\hat{\bm{f}}_{mn}^{\ddagger(k+1)},
    \label{eq:solNPSE}
\end{equation}
for $m=-M,\dots,M$ and $n=-N,\dots,N$. To solve~\eqref{eq:quasiNewton} or~\eqref{eq:solNPSE}, we mimic the NPSE solution procedure~\cite{Bertolotti_1991_Thesis,Chang_1993_PSE,Day_1999_Thesis}: we take the lower-upper (LU) decomposition of $[c^{(0)}A^{\ddagger}-\hat{L}_{mn}^{\ddagger}]$ to solve for $\hat{\bm{\phi}}_{m}^{\ddagger(k+1)}$, then we update the nonlinear term, and repeat until the residual is converged. We typically prefer solving~\eqref{eq:quasiNewton} over~\eqref{eq:solNPSE} because it reduces the number of iterations to convergence, without increasing the computational cost. Alternatively, if we accept a larger computational cost, then we can further reduce the number of iterations by solving~\eqref{eq:newton}.

The linear OWNS system~\eqref{02:eq:linSysFD} comprises $N_v+N_{\mathrm{aux}}=N_v+(2N_v+1)N_b$ equations, where $N_v$ scales with $N_y$, so that the computational cost to solve this system using a direct multifrontal solver (LU decomposition) scales as $\mathcal{O}(N_y^a N_b^a)$, where $a$ is a problem dependent coefficients that depend on the sparsity pattern.  Theoretically, $1 < a \leq 3$ and we typically observe $a \approx 1.5$ for 2D problems and $a \approx 2$ for 3D ones, see Towne et al.~\cite{Towne_2022_OWNS-P} for further details. The cost to integrate over $N_x$ stations is then $\mathcal{O}(N_xN_y^aN_b^a)$, while a global method entails a cost of $\mathcal{O}(N_x^aN_y^a)$, so that OWNS is more efficient for $N_b\ll N_x$. The nonlinear OWNS system~\eqref{02:eq:nlinSysFD} comprises $(M+1)\times(2N+1)\times(N_v+N_{\mathrm{aux}})$ equations, and can be solved using Newton's method~\eqref{eq:newton}, which entails a cost of $\mathcal{O}(M^aN^aN_y^aN_b^a)$. However, this cost can be reduced by instead solving~\eqref{eq:quasiNewton} or~\eqref{eq:solNPSE}, since we can perform the LU decomposition of $[c^{(0)}A^{\ddagger}-\hat{L}_{m}^\ddagger]$
separately for each Fourier mode, yielding a cost of $\mathcal{O}(MNN_y^aN_b^a)$. Therefore, the cost to integrate the NOWNS equations is $\mathcal{O}(N_xMNN_y^aN_b^a)$ using~\eqref{eq:quasiNewton} or \eqref{eq:solNPSE}, while it increases to $\mathcal{O}(N_xM^aN^aN_y^aN_b^a)$ for Newton's method~\eqref{eq:newton}, as compared to the cost $\mathcal{O}(N_x^aM^aN^aN_y^a)$ for nonlinear global methods.

Although~\eqref{eq:quasiNewton} and \eqref{eq:solNPSE} entail a lower computational cost, these methods fail for strong nonlinearities and we must instead employ Newton's method~\eqref{eq:newton}. In practice we implement a hybrid approach whereby we first solve~\eqref{eq:quasiNewton} to harness it's reduced computational cost, and then if more than $\mathcal{O}(100)$ iterations have elapsed, we switch to Newton's method to harness it's better convergence properties. We compare the performance of these three methods in appendix~\ref{sec:nonlinSolveStudy}.

To reduce the computational cost of Newton's method, we re-use the LU factors from the first iteration as a pre-conditioner for the Generalized Minimal Residual Method (GMRES), and we note that a similar procedure was performed for linear OWNS in Araya et al.~\cite{Araya_2022_FinnedCone}. We also tested the block-Jacobi relaxation method, which entails a cost of $\mathcal{O}(MNN_y^aN_b^a)$, and although we found this approach converged quickly in the early stages of the march, it failed as the nonlinearity grew stronger and we did not pursue relaxation methods further.

\subsubsection{Special treatment of the zero-frequency modes}

We have three options for obtaining a stable march for the zero-frequency modes, as depicted in table~\ref{tab:zFreq}. In the first approach, which mimics how the zero-frequency modes are handled by NPSE~\cite{Chang_1993_PSE,Day_1999_Thesis}, we neglect the streamwise presssure gradient, $\partial p_{0n}/\partial x$, and the streamwise diffusion terms, $\partial^2 q_{0n}/\partial x^2$, associated with the zero-frequency modes. However, in general we would prefer to avoid neglecting terms, and to instead parabolize these equations using the OWNS approach. In the second approach, we include the streamwise diffusion terms (but exclude the streamwise pressure gradient) associated with the zero-frequency modes, and parabolize the equations for all modes using OWNS. This approach yields a stable spatial march that agrees well with DNS, but offers no advantages over the first approach: for all of the cases examined in this paper, neglecting the streamwise diffusion terms associated with the zero-frequency modes does not change substantially the results of the NOWNS calculation. In the third approach, we include both the streamwise diffusion terms and the streamwise pressure gradient associated with the zero-frequency modes. Although the OWNS approach yields a stable spatial march, we have found that this approach produces inferior comparisons to DNS solutions from the literature, as discussed in appendix~\ref{sec:dp}.

In summary, including the streamwise pressure gradient for the zero-frequency modes worsens agreement with DNS, while including the streamwise diffusion terms for the zero-frequency modes has nearly no impact on the solution. Since neglecting these terms is more computationally efficient while still providing excellent agreement with DNS, we recommend that these terms be neglected.

\begin{table}
\centering
    \begin{tabular}{c|c|c|c|c}
         & Include $\partial p_{0n}/\partial x$? & Include $\partial^2 q_{0n}/\partial x^2$? & Parabolized using OWNS? & Agreement with DNS?\\
         \hline
         1 & No  & No  & No  & Yes\\
         2 & No  & Yes & Yes & Yes\\
         3 & Yes & Yes & Yes & No
    \end{tabular}
    \caption{Three approaches to parabolizing the zero-frequency modes}
    \label{tab:zFreq}
\end{table}

\subsubsection{Spanwise symmetry} \label{sec:sym}

If the disturbances are symmetric, then we can enforce a symmetry condition to reduce the number of equations from $(M+1)\times(2N+1)$ to $(M+1)\times(N+1)$. All variables have even-symmetry ($\hat{\nu}_{m,-n}'=\hat{\nu}_{m,n}'$, $\hat{u}_{m,-n}'=\hat{u}_{m,n}'$, $\hat{v}_{m,-n}'=\hat{v}_{m,n}'$, $\hat{p}_{m,-n}'=\hat{p}_{m,n}'$), with the exception of the $w$-velocity which has odd-symmetry ($\hat{w}_{m,-n}'=-\hat{w}_{m,n}'$).

\subsubsection{Boundary conditions}
 
At the wall, we impose no-slip isothermal boundary conditions ($u'=v'=w'=T'=0$) and solve for the specific volume, $\nu'$, using the (nonlinear) continuity equation. At the far-field boundary, we impose 1D (in $y$) inviscid Thompson characteristic boundary conditions to minimize spurious numerical reflections~\cite{Thompson_1987_BC}, which we implement using the linearized boundary-layer flow equations.
 
Some previous work on NPSE has used similar characteristic far-field boundary conditions~\cite{Day_1999_Thesis}. Chang et al.~\cite{Chang_1993_PSE} used the far-field boundary condition $\hat{\bm{q}}_{mn}'(y_{\max})=0$ for $m, n \ne 0$. As the boundary layer must be allowed to grow in the wall-normal direction (due to nonlinear interactions), they used $\partial\hat{v}_{00}'/\partial y=0$ at $y_{\max}$ for the mean-flow distortion (MFD). The characteristic far-field boundary conditions are advantageous because they allow us to use the same boundary conditions for all Fouriers modes, instead of handling the MFD as a separate case.

\subsubsection{Effects of the mean-flow distortion}

In NOWNS, the disturbances interact to excite the MFD, so that the corrected mean flow, $\bar{\bm{q}}+\hat{\bm{q}}_{00}$, differs from the baseflow, $\bar{\bm{q}}$. We have experimented with linearizing about both the baseflow and the corrected mean flow and found that it does not have a large impact on the NOWNS calculation, as discussed in appendix~\ref{sec:meanFlowStudy}. Linearizing about the corrected mean flow increases the computational cost of NOWNS because the projection operators change between iterations, since the MFD changes, so the LU factorization must be updated. On the other hand, linearizing about the baseflow allows us to use the same LU factorization at each iteration because the baseflow is not affected by changes in the MFD. Since it is more computationally efficient to linearize only about the baseflow, we choose this approach moving forward.

\subsubsection{Recursion parameters}

The choice of recursion parameters is described in Sleeman et al.~\cite{Sleeman_2024_NOWNS-Aviation}, and matches the recursion parameters used by Rigas et al.~\cite{Rigas_2017_OWNS_BL}, which are based on the recursion parameters originally developed by Towne and Colonius~\cite{Towne_2015_OWNS-O}. We briefly discuss our strategy for selecting the recursion parameters when $\omega=0$ in appendix~\ref{app:recursions}.

\section{Validation}\label{sec:validation}

We validate NOWNS by applying it to 2D and 3D Blasius boundary layer flows for which there are existing DNS and NPSE results in the literature. We choose a Mach number of $\mathrm{Ma}_{\infty}=0.1$ to study flows near the incompressible limit.
In what follows, we use the following dimensionless quantities:
\[
\mathrm{Re}_x = \frac{U_\infty^* x^*}{\nu_\infty^* \mu_\infty^*},\quad
y=\frac{y^*}{\delta_0^*},\quad
F=\frac{\omega^{*}\nu_\infty^*\mu_\infty^*}{U_\infty^{*2}},\quad
b=\frac{\beta^*\nu_\infty^*\mu_\infty^*}{U_\infty^*},
\]
where $\mathrm{Re}_x$ is the streamwise coordinate, $F$ is the temporal frequency, and $b$ is the spanwise wavenumber. We refer to modes according to their temporal frequency and their spanwise wave number as $(m,n)$, where $m$ refers to the frequency $\omega_m=m\omega$ and $n$ refers to the spanwise wave number $\beta_n=n\beta$. To be consistent with previous literature, we measure the amplitude of disturbances as
\begin{align}
    u^{\prime(m,n)}_{\max}(x) = c_{m,n} \ \max_{y}
    |u_{m,n}'(x,y)|, \qquad c_{m,n} = \begin{cases}
     1 & \mbox{$m=n=0$}, \\
     \sqrt{2} & \mbox{$m=0, n \neq 0$; $n=0, m \neq 0$}, \\
    2 & \mbox{otherwise}. \end{cases}
\end{align}

\subsection{2D evolution of a Tollmien-Schlichting wave}\label{sec:results-2D}

We consider the test case developed by Bertolotti et al.~\cite{Bertolotti_1992_PSE} which has been widely used in the literature as a validation case for NPSE~\cite{Joslin_1992_DNS,Joslin_1993_DNS,Paredes_2015_PSE3D}. This case examines the evolution of a Tollmien-Schlichting  (TS) wave excited at the inlet at a frequency $F = 86\times10^{-6}$  and amplitude  $u^{\prime(1)}_{\max}(x_0)=0.25\%$. All other Fourier components initially have zero-amplitude and are generated through nonlinear interactions with the TS wave. The grid extends over the domain $\mathrm{Re}_x\in[1.6\times10^5,10^6]$ and $y\in[0,75]$ with $4000$ stations evenly spaced in $x$ and 150 grid points in $y$, with the majority of the grid points clustered towards the wall, while the Fourier series is truncated at $M=5$ temporal modes.

\begin{figure}
    \centering
    \includegraphics[width=0.5\columnwidth]{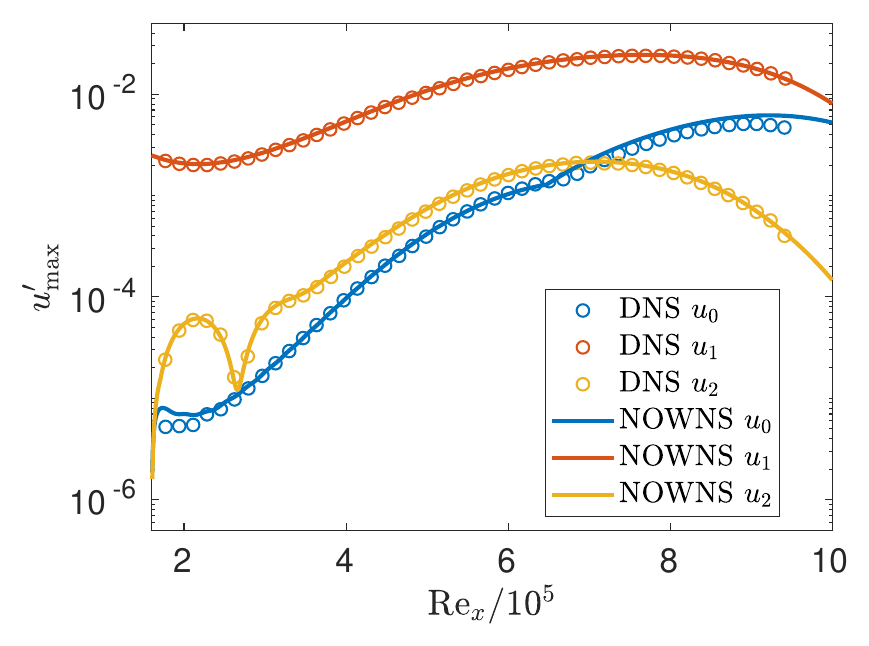}
    \caption{Amplitude of $u'$ v.s.\ streamwise coordinate, $\mathrm{Re}_x$, for 2D evolution of TS wave with $u_{\max}'^{(1)}(x_0)=0.25\%$.}
    \label{fig:0p25_chang-7b}
\end{figure}

\begin{figure}
    \centering
    \begin{subfigure}[b]{0.3\textwidth}
        \centering
        \includegraphics[width=\textwidth]{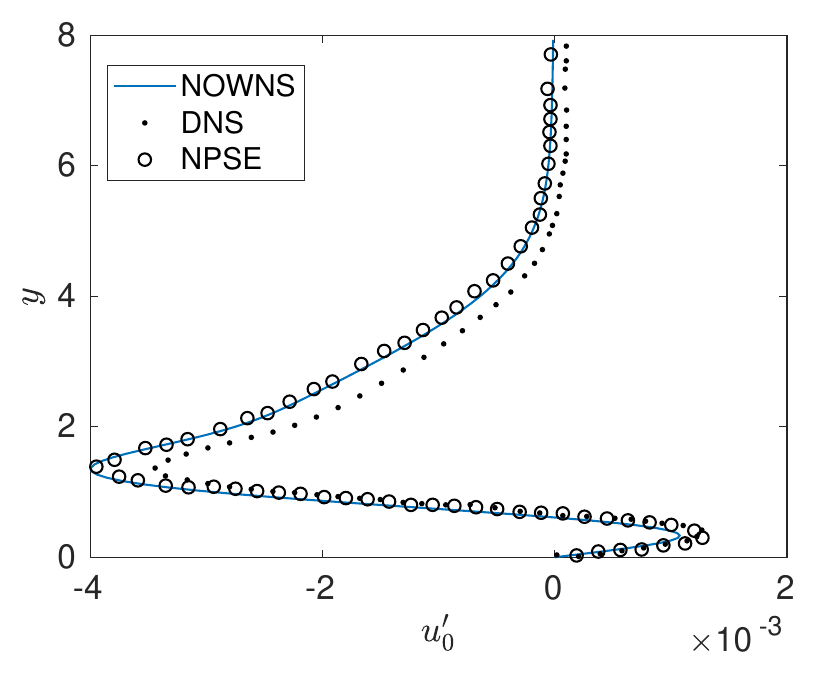}
        \caption{$u$-velocity for MFD}
        \label{fig:0p25-u0}
    \end{subfigure}
    \begin{subfigure}[b]{0.3\textwidth}
        \centering
        \includegraphics[width=\textwidth]{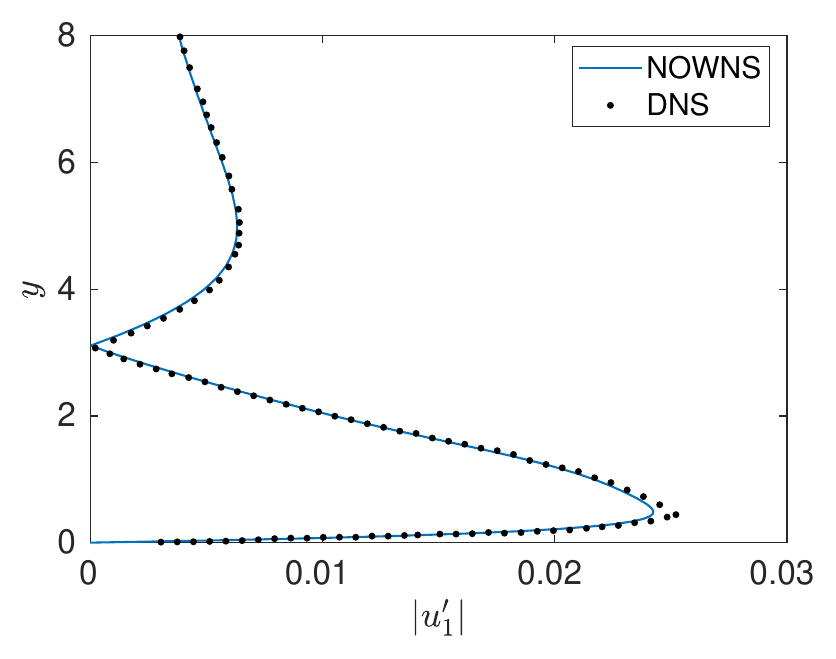}
        \caption{$u$-velocity for TS wave}
        \label{fig:0p25-u1}
    \end{subfigure}     
    
    \bigskip
    \begin{subfigure}[b]{0.3\textwidth}
        \centering
        \includegraphics[width=\textwidth]{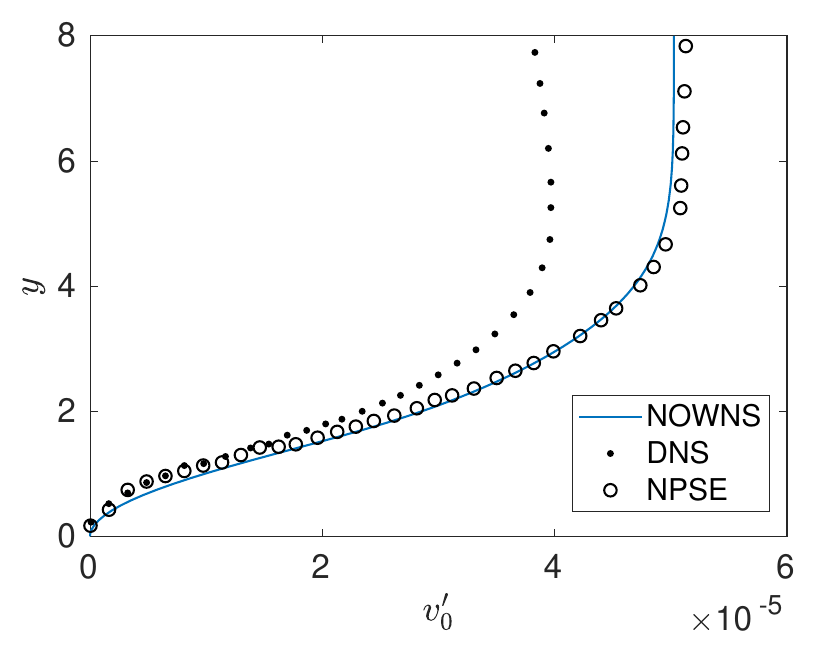}
        \caption{$v$-velocity for MFD}
        \label{fig:0p25-v0}
     \end{subfigure}
     \begin{subfigure}[b]{0.3\textwidth}
        \centering
        \includegraphics[width=\textwidth]{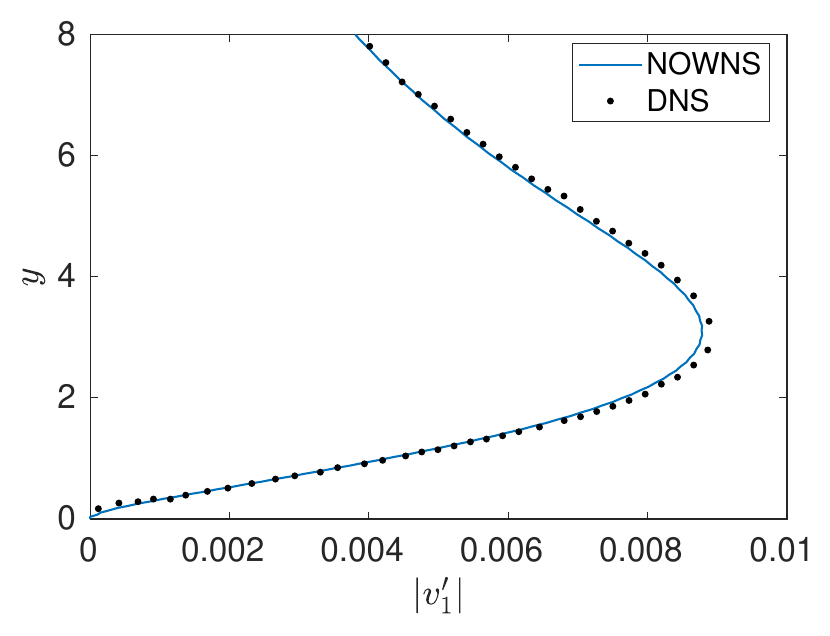}
        \caption{$v$-velocity for TS wave}
        \label{fig:0p25-v1}
     \end{subfigure}
     \caption{$u$-and $v$-velocity profiles at streamwise coordinate $\mathrm{Re}_x=7.80\times10^5$ for 2D evolution of TS wave.}
     \label{fig:0p25_joslin-2}
\end{figure}

Figure~\ref{fig:0p25_chang-7b} compares NOWNS to DNS for $u^{\prime(m,n)}_{\max}(x)$, while the $u$- and $v$-velocity profiles for the MFD and TS waves are shown in figure~\ref{fig:0p25_joslin-2}. Excellent agreement is obtained; the discrepancy for the MFD of $v$ can be attributed to the Dirichlet boundary conditions used in the DNS.

\subsection{3D oblique-wave breakdown}\label{sec:valid-OW}

Next we consider the oblique-wave breakdown case studied by Joslin et al.~\cite{Joslin_1993_DNS}, where transition is initiated by two oblique waves with opposite wave angle. They defined two cases--with small and large initial amplitude. In this section, for validation purposes, we consider the small amplitude case for which NPSE was previously successful.
Oblique-wave breakdown has also been studied using both experiment and spatial DNS by Berlin et al.~\cite{Berlin_1999_Oblique}, while it was studied using NPSE for compressible flows by Chang and Malik~\cite{Chang_1994_Oblique}. We further note that whereas fundamental and subharmonic transition can be studied using Herbert's secondary stability theory~\cite{Herbert_1988_Secondary}, no such theory exists for oblique-wave breakdown, so that either experiment or numerical simulation is necessary to study this transition scenario~\cite{Joslin_1993_DNS,Berlin_1999_Oblique}.

\begin{figure}
    \centering
    \includegraphics[width=0.6\columnwidth]{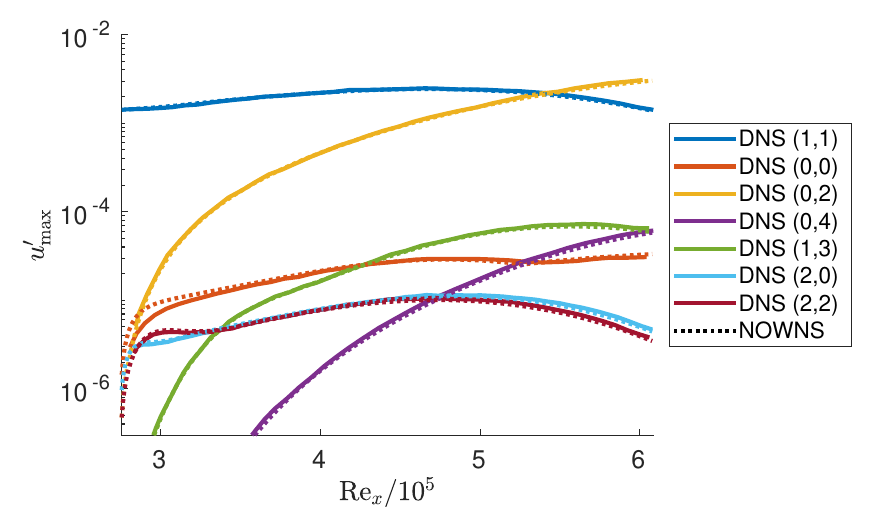}
    \caption{Amplitude of $u'$ v.s.\ streamwise coordinate, $\mathrm{Re}_x$, for the small-amplitude oblique-wave breakdown case at frequency $F=86\times10^{-6}$, spanwise wavenumber $b=2/9\times10^{-3}$, with initial amplitude of $u_{\max}^{\prime(1,1)}(x_0)=\sqrt{2}\times10^{-3}$.}
    \label{fig:joslin_1993_12}
\end{figure}

\begin{figure}
     \centering
     \begin{subfigure}[b]{0.3\textwidth}
         \centering
         \includegraphics[width=\textwidth]{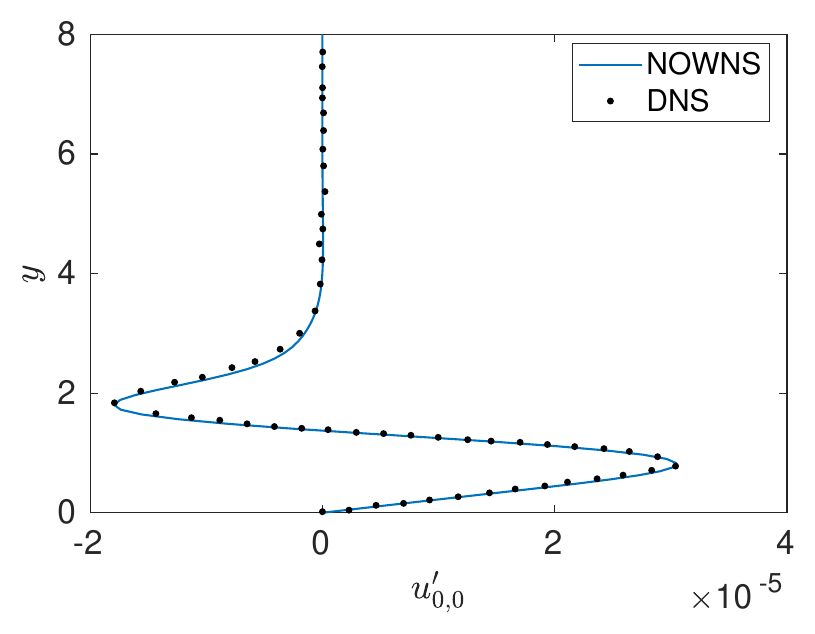}
         \caption{MFD}
         \label{fig:joslin_1993_13_0c0}
     \end{subfigure}
     \begin{subfigure}[b]{0.3\textwidth}
         \centering
         \includegraphics[width=\textwidth]{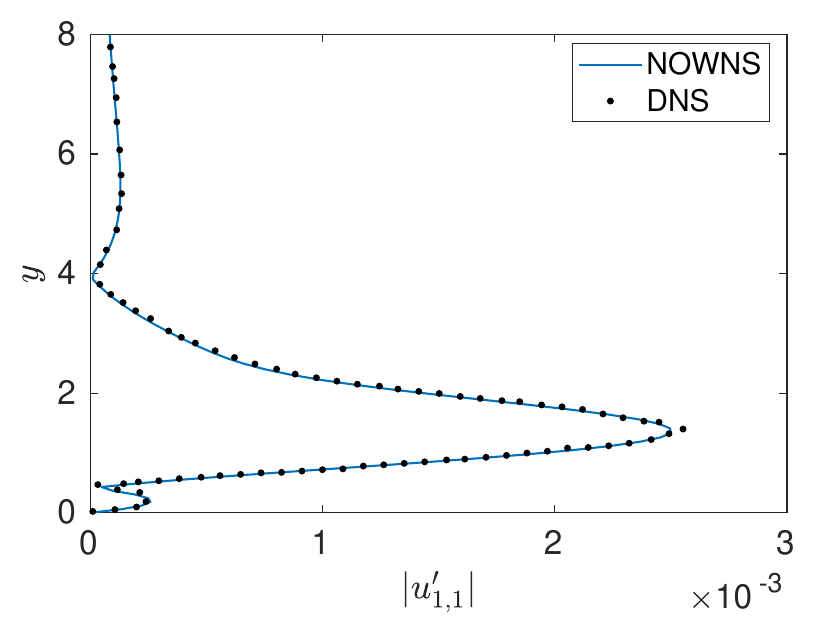}
         \caption{Oblique mode}
         \label{fig:joslin_1993_13_1c1}
     \end{subfigure}
     
     \bigskip
     \begin{subfigure}[b]{0.3\textwidth}
         \centering
         \includegraphics[width=\textwidth]{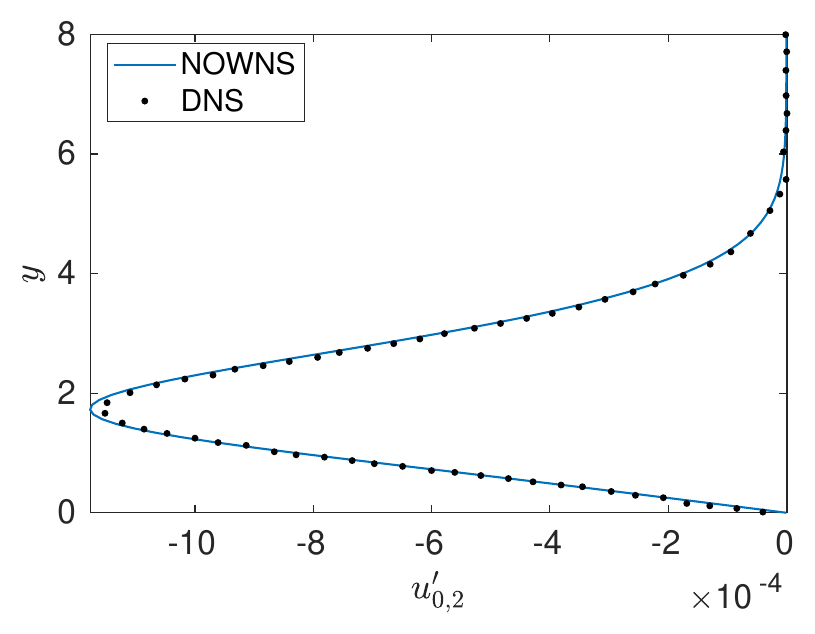}
         \caption{Vortex mode}
         \label{fig:joslin_1993_13_0c2}
     \end{subfigure}
     \begin{subfigure}[b]{0.3\textwidth}
         \centering
         \includegraphics[width=\textwidth]{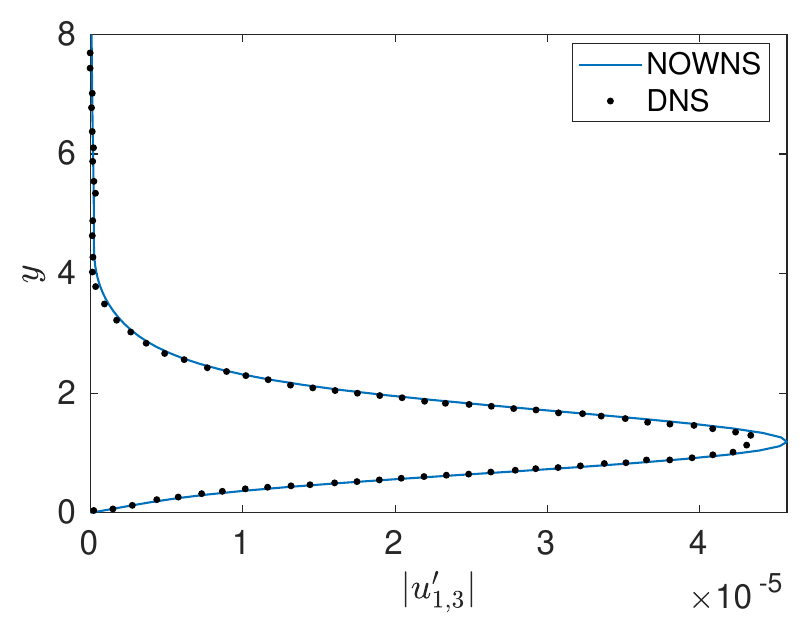}
         \caption{First harmonic in $z$}
         \label{fig:joslin_1993_13_1c3}
     \end{subfigure}
     
        \caption{$u$-velocity profiles at streamwise coordinate $\mathrm{Re}_x=4.69\times10^5$ for small-amplitude oblique-wave breakdown.}
        \label{fig:joslin_1993_13_a}
\end{figure}

The oblique waves have amplitude $u_{\max}^{\prime(1,1)}(x_0)=\sqrt{2}\times10^{-3}$ at the inlet at a frequency $F=86\times10^{-6}$ and spanwise wavenumber $b=2/9\times10^{-3}$, while the grid extends over the domain $\mathrm{Re}_x\in[2.74\times10^5,6.08\times10^5]$ and $y\in[0,75]$ with $2000$ stations evenly spaced in $x$ and 150 grid points in $y$, while the Fourier series is truncated at $M=3$ and $N=4$. Figure~\ref{fig:joslin_1993_12} compares NOWNS to DNS and NPSE for $u_{\max}^{\prime(m,n)}(x)$, while figure~\ref{fig:joslin_1993_13_a} compares the $u$-velocity profiles at $\mathrm{Re}_x=4.69\times10^5$; we see that we have excellent agreement for between the DNS and NOWNS results for all for modes.

\section{A Comparison of NOWNS to NPSE}

In this section, we apply NOWNS in three scenarios where NPSE is known to break down, to demonstrate that NOWNS can succeed where NPSE fails.

\subsection{High amplitude evolution of TS wave}\label{sec:largeAmplitude}

It is well-known that NPSE can fail for sufficiently strong nonlinearities~\cite{Day_1999_Thesis,Towne_2019_PSE}. In Sleeman et al.~\cite{Sleeman_2024_NOWNS-Aviation}, we applied NOWNS to the high amplitude oblique wave-breakdown case studied in Joslin et al.~\cite{Joslin_1993_DNS} and demonstrated that although their NPSE calculation fails, the NOWNS calculation does not. However, the DNS calculation is under-resolved, and the NOWNS calculation does not match the DNS near the end of the domain, which does not demonstrate that NOWNS can be useful past the point where NPSE fails. Therefore, we consider a 2D test case from Scholten et al.~\cite{Scholten_2024_PSE}, whereby we modify the validation case discussed in section~\ref{sec:results-2D}, such that the we march over the domain $\mathrm{Re}_x\in[1.6\times10^5,8.59\times10^5]$ with 2663 stations evenly spaced in $x$, while the wall-normal domain remains the same as before, and the Fourier series truncate at $M=10$ temporal modes. Scholten et al.~\cite{Scholten_2024_PSE} compute a high-fidelity solution using a harmonic Navier-Stokes equation (HNSE) solver, and demonstrate that NPSE agrees with HNSE with an inlet amplitude of $u_{\max}'^{(1)}(x_0)=0.4\%$, but that the NPSE march fails. In figure~\cite{Scholten_2024_PSE}, we plot the NOWNS amplitudes against those from HNSE and demonstrate that NOWNS marches farther than NPSE, while still being accurate, but that it too fails. This highlights that NOWNS can march farther than NPSE, while still being accurate, and that the failure of NPSE may not be related to its minimum step-size requirement or the regularization techniques it uses to suppress the upstream-going waves.

\begin{figure}
    \centering
    \includegraphics[width=0.6\columnwidth]{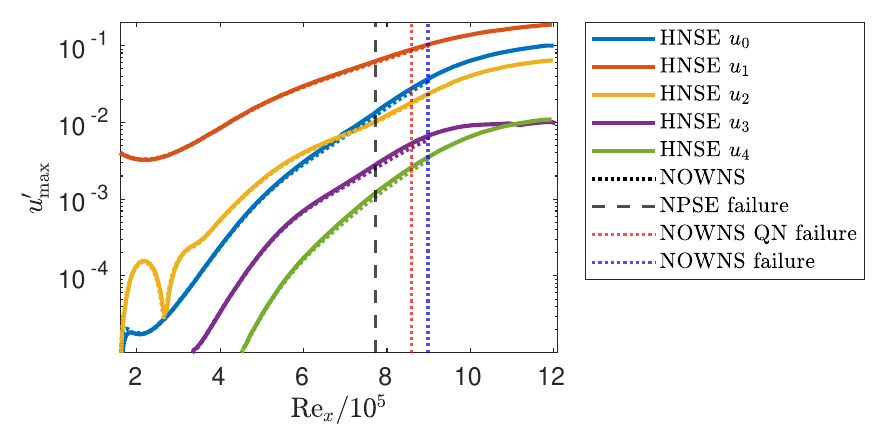}
    \caption{Amplitude of $u'$ v.s.\ streamwise coordinate, $\mathrm{Re}_x$, for 2D evolution of TS wave with $u_{\max}'^{(1)}(x_0)=0.4\%$.}
    \label{fig:0p4_Scholten_Amplitudes}
\end{figure}

As discussed in~\ref{subsubsec:compCost}, the quasi-Newton method is more computationally efficient than the full Newton's method, but takes more iterations to converge. In addition, the full Newton's method is more likely to converge for stronger nonlinearities. In figure~\ref{fig:0p4_Scholten_Amplitudes}, we indicate where each of the the quasi-Newton and full Newton methods fail. Even without employing the full Newton's method, NOWNS is able to march farther downstream than NPSE.

\subsection{Low amplitude oblique-wave breakdown with random noise}\label{sec:randNoise}

\begin{figure}
    \centering
    \begin{subfigure}[b]{0.3\textwidth}
        \includegraphics[width=\textwidth]{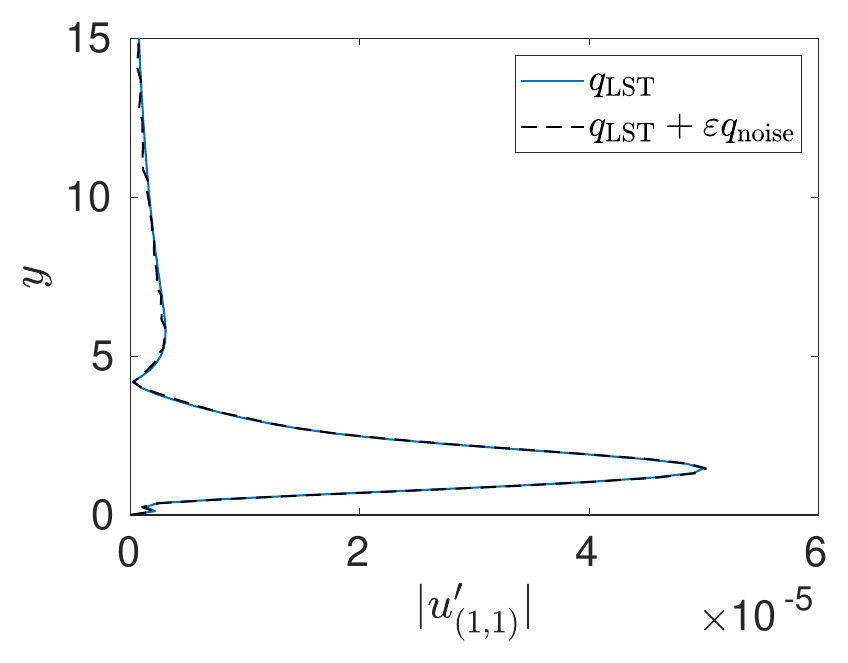}
    \end{subfigure}
    \begin{subfigure}[b]{0.3\textwidth}
        \includegraphics[width=\textwidth]{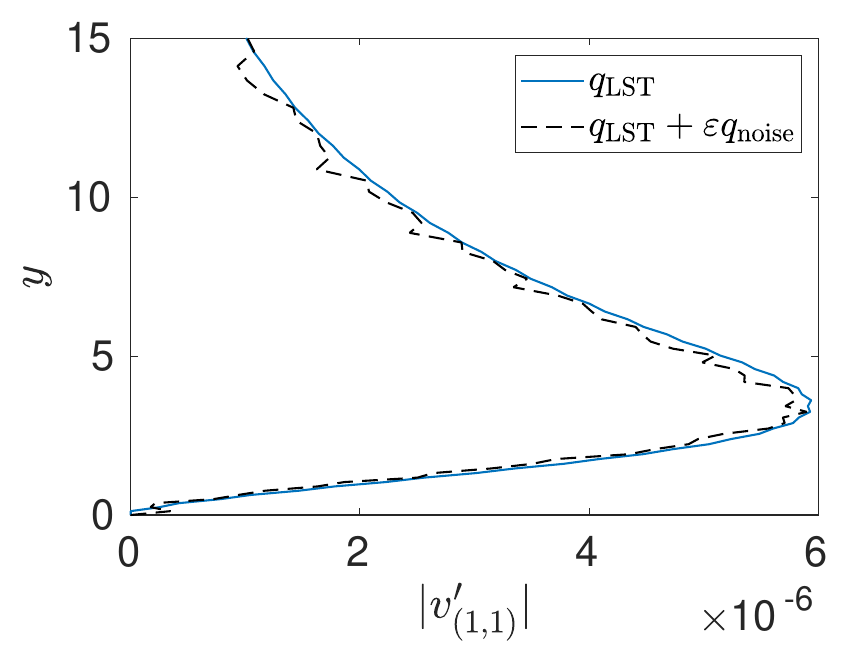}
    \end{subfigure}
    
    \bigskip
    \begin{subfigure}[b]{0.3\textwidth}
        \includegraphics[width=\textwidth]{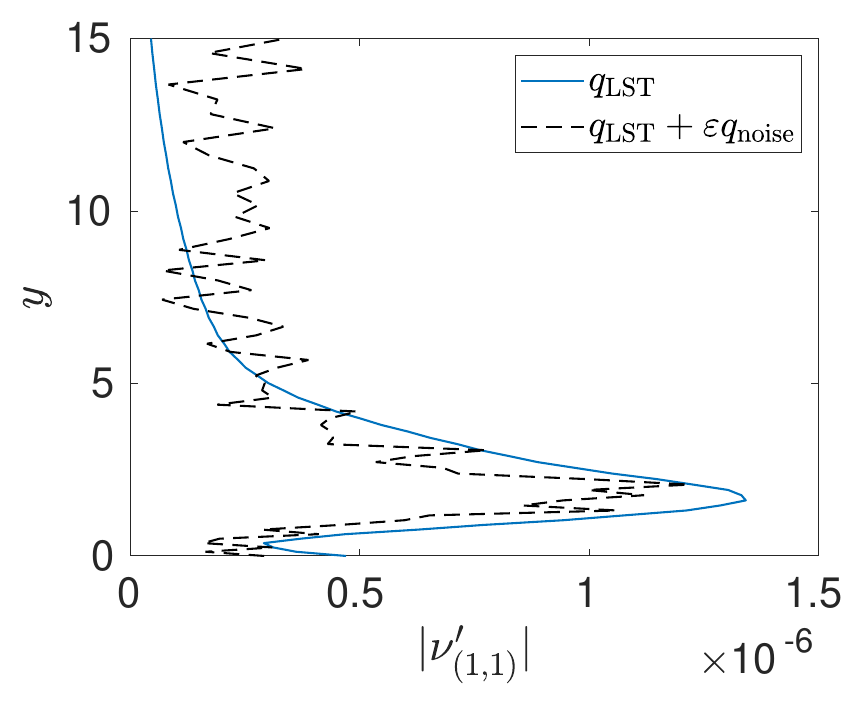}
    \end{subfigure}
    \begin{subfigure}[b]{0.3\textwidth}
        \includegraphics[width=\textwidth]{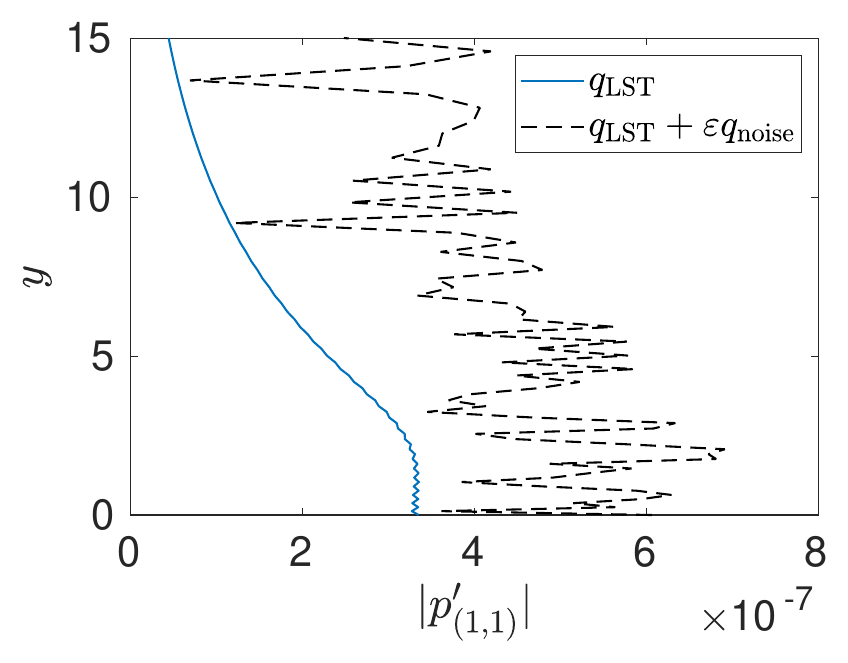}
    \end{subfigure}    
    \caption{Oblique-wave inlet boundary condition with and without random noise.}
    \label{fig:rand_profiles_1}
\end{figure}

Here we demonstrate for the low amplitude oblique wave breakdown case of section~\ref{sec:valid-OW} that NOWNS is robust to random noise applied to the inlet boundary condition. Given the eigenfunction from the locally parallel linear stability theory, $\bm{q}_{\mathrm{LST}}$, we add random noise, $\bm{q}_{\mathrm{noise}}$, to obtain the inlet condition $\bm{q}_{\mathrm{LST}}+\varepsilon \tilde{\bm{q}}_{\mathrm{noise}}$. We choose complex random noise such that
\[
\tilde{\bm{q}}_{\mathrm{noise}}=\tilde{\bm{q}}_{\mathrm{noise},r}+i \tilde{\bm{q}}_{\mathrm{noise},i},
\quad \tilde{\bm{q}}_{\mathrm{noise},r},\tilde{\bm{q}}_{\mathrm{noise},i}\sim \mathcal{U}_{[0,1]},
\]
where $\mathcal{U}_{[a,b]}$ represents the uniform distribution over the interval $[a,b]$. We then normalize the noise to obtain $q_{\mathrm{noise}}$, such that the maximum amplitude of the $u$-velocity noise is equal to the free-stream $u$-velocity, $U_\infty$. We recall that the amplitude of $\bm{q}_{\mathrm{LST}}$ is $\sqrt{2}\times10^{-3}$, and we choose $\varepsilon=\sqrt{2}\times10^{-5}$. In figure~\ref{fig:rand_profiles_1}, we plot the profile of the $u$- and $v$-velocities, as well as the thermodynamic variables. We see that the random noise has a relatively small effect on the $u$-velocity, and a slightly more pronounced effect on the $v$-velocity, while it has a larger impact on the thermodynamic variables. Like the $u$-velocity profile, the random noise has a relatively small effect on the $w$-velocity profile, so we omit this plot.

\begin{figure}
    \centering
    \includegraphics[width=0.6\columnwidth]{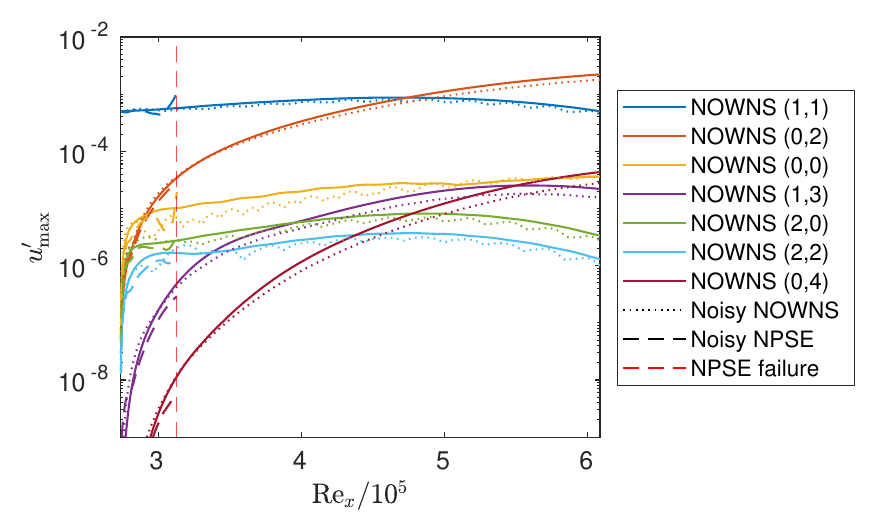}
    \caption{Amplitude of $u'$ v.s.\ streamwise coordinate, $\mathrm{Re}_x$, for low amplitude oblique-wave breakdown with and without random noise.}
    \label{fig:rand_amplitudes}
\end{figure}

\begin{figure}
    \centering
    \begin{subfigure}[b]{0.45\textwidth}
        \includegraphics[width=0.95\columnwidth]{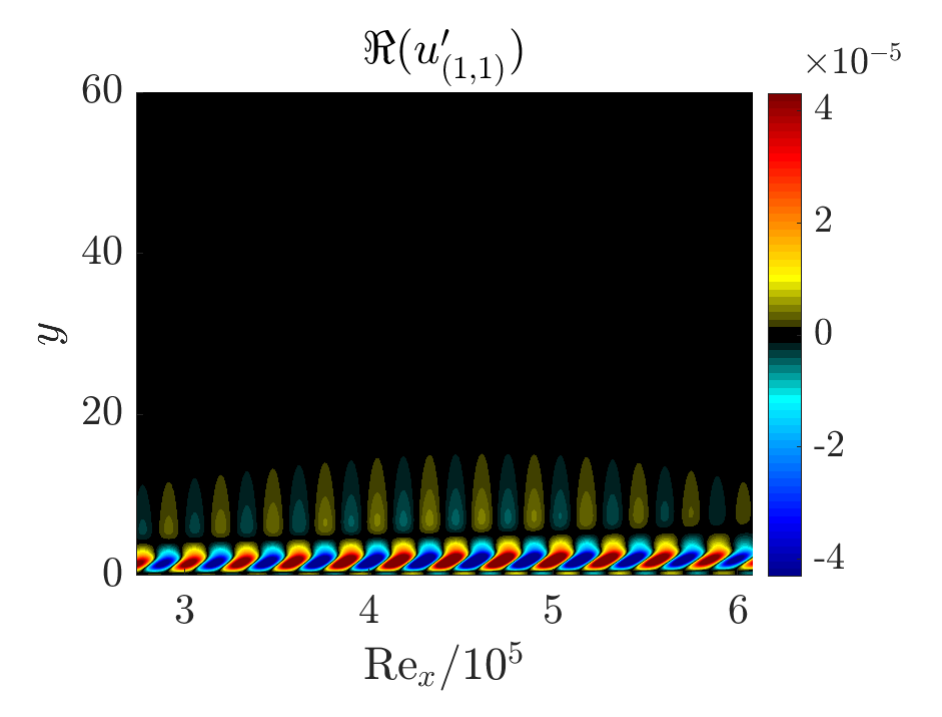}
    \end{subfigure}
    \begin{subfigure}[b]{0.45\textwidth}
        \includegraphics[width=0.95\columnwidth]{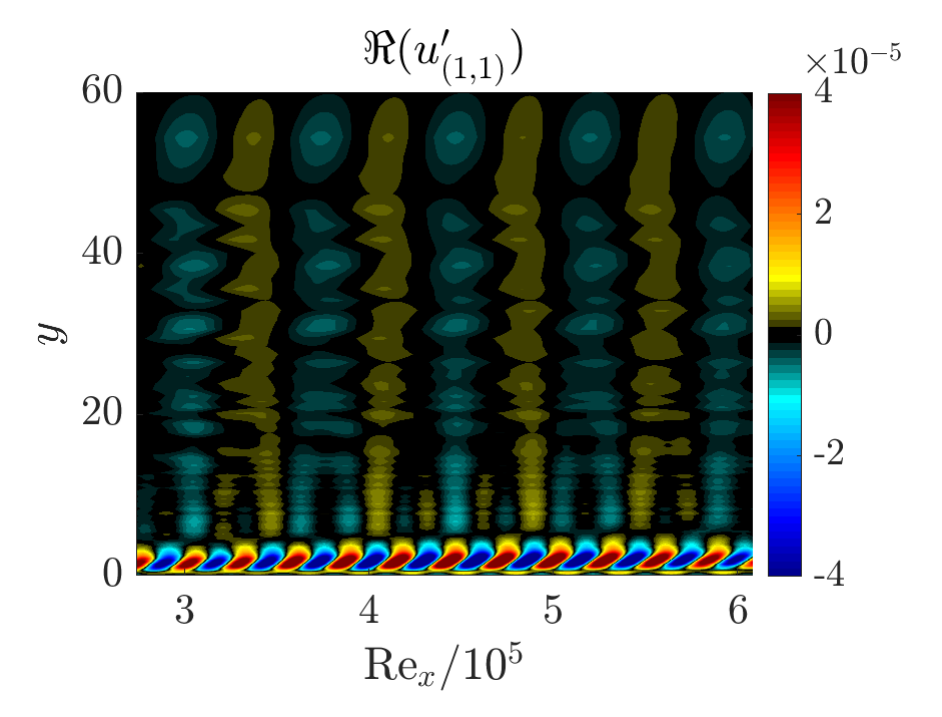}
    \end{subfigure}

    \bigskip
    \begin{subfigure}[b]{0.45\textwidth}
        \includegraphics[width=0.95\columnwidth]{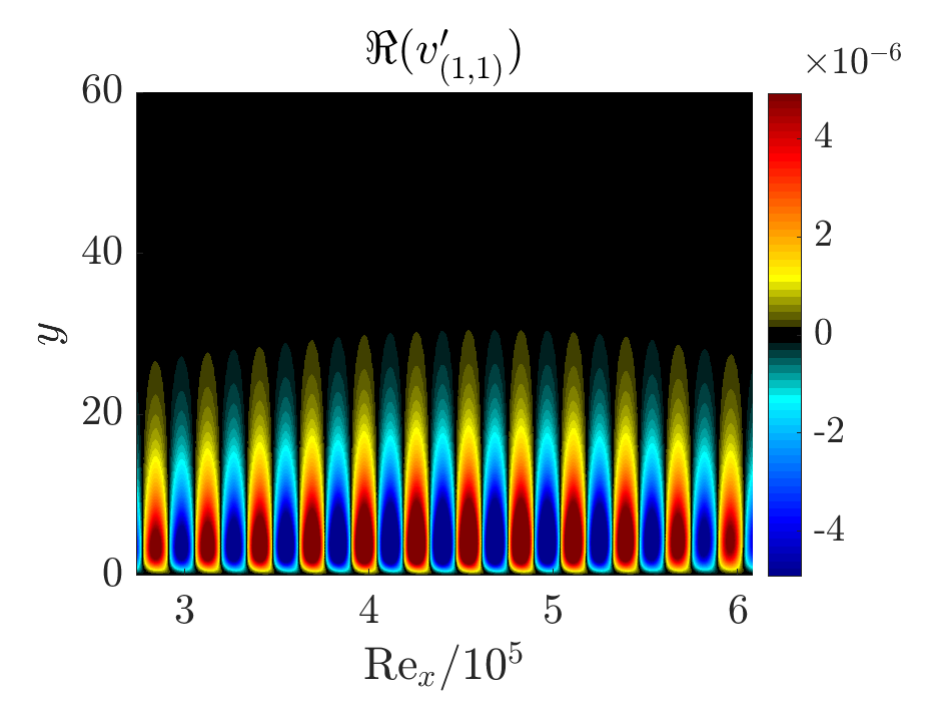}
    \end{subfigure}
    \begin{subfigure}[b]{0.45\textwidth}
        \includegraphics[width=0.95\columnwidth]{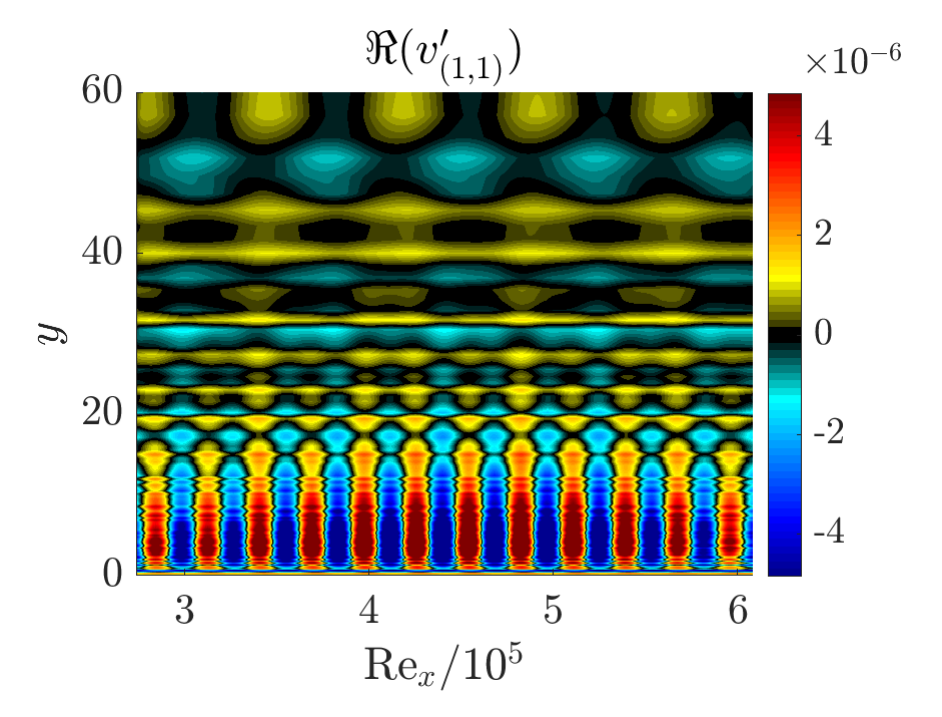}
    \end{subfigure}    
    \caption{Contour plots of the $u$- and $v$ velocities of the oblique wave, with and without random noise.}
    \label{fig:rand_contour}
\end{figure}

The NOWNS march succeeds even if we introduce large disturbances to the velocity fields, but the $u'$ amplitudes differ substantially from the noise-less case due to the large perturbations. Therefore, we instead introduce relatively large random disturbances to the thermodynamic variables, and relatively small ones to the velocity field. In figure~\ref{fig:rand_amplitudes}, we see that the amplitudes predicted by NOWNS for the noisy inlet condition agree closely with those without noise. We also see that although NPSE initially is able to accurately predict the evolution of the $u$-velocity amplitudes, it eventually becomes inaccurate (especially for $u_{(0,0)}'$ and $u_{(1,1)}'$) before failing. We further note that increasing the resolution of the NPSE calculation in Fourier space does not help its convergence, so its failure is not because it is under-resolved. We plot the contours of the real part of the $u$- and $v$-velocities of the oblique wave, with and without noise in figure~\ref{fig:rand_contour}. We see that despite the noisy inlet condition, NOWNS evolves the $u$-velocity of the oblique wave such that it matches closely the case without noise. On the other hand, the $v$-velocity is more affected by the numerical noise, yet we still obtain good qualitative agreement. The other modes (e.g., the vortex mode) are evolved accurately by the NOWNS calculation with noisy inlet condition, and the contour plots with and without the numerical noise are indistinguishable from each other, and so are not plotted here.

\subsection{Blowing/suction strip}~\label{sec:kType}

Blowing/suction strips are frequently used to study laminar-turbulent boundary layer transition in low-speed boundary layer flows~\cite{Fasel_1990_Subharmonic,Rist_1995_Subharmonic,Sayadi_2013_DNS,Huai_1997_LES,Rigas_2021_NonlinearIO}. Here, we introduce disturbances by specifying a non-zero wall-normal velocity such that $v(y=0)=f(x,z,t)$, for some function $f(x,z,t)$ that is periodic in $t$ and $z$. NPSE cannot be used to simulate disturbances introduced using blowing/suction strips because they are non-modal, demonstrating an advantage of NOWNS. However, we note that Herbert's second stability theory yields an inlet boundary condition so that NPSE (and NOWNS) can be used to study fundamental (K-type) and subharmonic (H-type) transition~\cite{Herbert_1988_Secondary,Herbert_1997_PSE}.

Rist and Fasel~\cite{Rist_1995_Subharmonic} used DNS with a blowing/suction strip to study K-type transition, while similar studies were performed by Sayadi et al.~\cite{Sayadi_2013_DNS}, and by Rigas et al.~\cite{Rigas_2021_NonlinearIO} using a harmonic balance method (HBM). The blowing/suction strip is given by
\begin{equation}
    f(x,z,t)=5\times10^{-3}\sin(\omega t) v_a(x) + 1.3\times10^{-4}\cos(\beta z)v_s(x),
\end{equation}
where
\begin{subequations}
    \begin{equation}
        v_a(\mathrm{Re}_x)=\left\{
        \begin{array}{cc}
             0, &  \mathrm{Re}_x \leq \mathrm{Re}_x(x_1)\\
             15.1875\xi^{5}-35.4375\xi^{4}+20.25\xi^{3}, & \mathrm{Re}_x(x_1)<\mathrm{Re}_x\leq \mathrm{Re}_x(x_m)\\
             -v_a\big(2\mathrm{Re}_x(x_m)-\mathrm{Re}_x\big), & \mathrm{Re}_x(x_m)<\mathrm{Re}_x\leq \mathrm{Re}_x(x_2)\\
             0, & \mathrm{Re}_x(x_2) < \mathrm{Re}_x
        \end{array}
        \right.
    \end{equation}
    \begin{equation}
        v_s(\mathrm{Re}_x)=\left\{
        \begin{array}{cc}
             0, &  R \leq \mathrm{Re}_x(x_1)\\
             -3\xi^{4}+4\xi^{3}, & \mathrm{Re}_x(x_1)<\mathrm{Re}_x\leq\mathrm{Re}_x(x_m)\\
             v_s\big(2\mathrm{Re}_x(x_m)-\mathrm{Re}_x\big), & \mathrm{Re}_x(x_m)<\mathrm{Re}_x\leq \mathrm{Re}_x(x_2)\\
             0, & \mathrm{Re}_x(x_2) < \mathrm{Re}_x
        \end{array}
        \right.
    \end{equation}
\end{subequations}
for $\mathrm{Re}_x(x_1)=1.3438\times10^5$, $\mathrm{Re}_x(x_2)=1.5532\times10^5$, $x_m=(x_1+x_2)/2$, and $\xi=(\mathrm{Re}_x-\mathrm{Re}_x(x_1))/(\mathrm{Re}_x(x_m)-\mathrm{Re}_x(x_1))$. We choose $F=110\times10^{-6}$ and $b=0.423\times10^{-3}$ with $M=N=4$, while the grid extends over $\mathrm{Re}_x\in[1.33956\times10^5,2.86\times10^5]$ and $y\in[0,60]$, with 1413 stations in $x$ and 100 grid points in $y$.

\begin{figure}
    \centering
    \includegraphics[width=0.8\columnwidth]{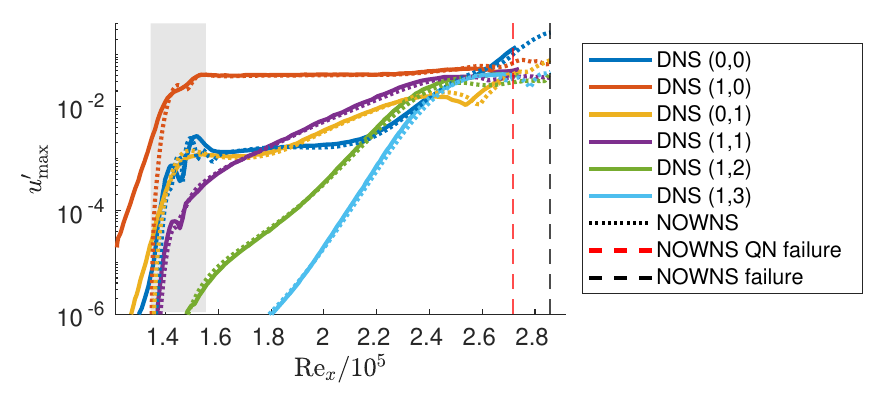}
    \caption{Amplitude of $u'$ v.s.\ streamwise coordinate, $\mathrm{Re}_x$, for K-type transition}
    \label{fig:ff_amplitudes}
\end{figure}

\begin{figure}
    \centering
    \includegraphics[width=0.8\columnwidth]{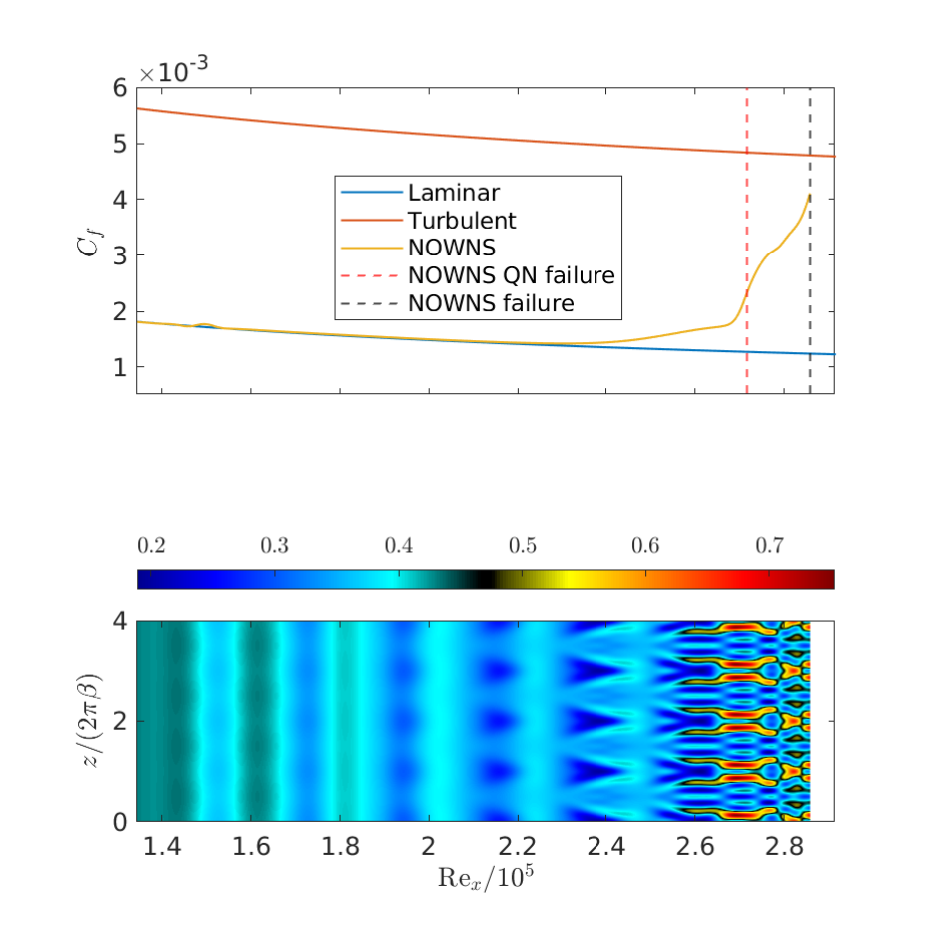}
    \caption{Skin-friction coefficient (top) and $u$-velocity contours (bottom) as a function of streamwise coordinate. The skin-friction coefficient rises rapidly following the onset of transition, before the march fails, while the $u$-velocity contour plot shows the aligned $\Lambda$-vortex structure characteristic of K-type transition.}
    \label{fig:Ktype_CfDelta}
\end{figure}

Figure~\ref{fig:ff_amplitudes} shows excellent agreement between the $u'$ amplitudes of NOWNS and the DNS of Rist and Fasel~\cite{Rist_1995_Subharmonic}. We note that in the early stages of the march there is disagreement between the DNS and NOWNS calculations because the blowing/suction strip causes upstream effects that NOWNS neglects by construction. However, these disturbances are convective in nature, and the amplitudes predicted by NOWNS rapidly converge to those predicted by DNS as the march progresses downstream. We note that the NOWNS calculation marches past the last station for which DNS $u$-velocity amplitudes are provided by Rist and Fasel, so we continue plotting the NOWNS amplitudes without DNS data. Figure~\ref{fig:Ktype_CfDelta} shows that the skin-friction rises rapidly following the onset of laminar-turbulent transition, before the NOWNS march fails. This figure also shows the contours of the $u$-velocity, demonstrating the align-lambda vortex structure that is characteristic of K-type transition. We further note that the quasi-Newton method approach fails following the initial rise of the skin-friction coefficient, and that the full Newton's method becomes necessary following the onset of transition. Lozano-Duran et al.~\cite{LozanoDuran_2018_PSE-LES} have combined NPSE with wall-resolved LES (WRLES), whereby NPSE is used to simulate the early stages of the march and is used as an inlet boundary condition for the WRLES calculation. This reduces the computational cost of WRLES, because its streamwise domain is smaller. A similar strategy using NOWNS with WRLES could be pursued in the future.

\section{Conclusion}

We have extended the OWNS approach, a fast marching procedure previously developed for solving linear flow disturbance equations, to support nonlinear interactions. We have demonstrated that it is effective for 2D and 3D disturbances in a wall-bounded flow by comparing against the spatial DNS and NPSE studies of Joslin et al.~\cite{Joslin_1993_DNS}. We also highlighted that although NOWNS is more expensive than NPSE in cases where NPSE succeeds, NOWNS is successful in cases where NPSE fails, which justifies its higher cost. In particular, we have shown that NOWNS can handle stronger nonlinearities and continues marching past the point where NPSE fails, and that NOWNS is robust to numerical noise. We further demonstrated that NOWNS accurately evolves non-modal instabilities introduced using a blowing/suction strip, which cannot be done using NPSE. Future work will extend the NOWNS procedure to support high-speed flows and will concentrate on studying optimal transition mechanisms, seeking to apply NOWNS in cases where spatial DNS and other global methods are too computationally expensive to be feasible.

\appendix

\section{Navier-Stokes Equations}\label{app:nse}

Given the non-dimensional Navier-Stokes equations~\ref{eq:nonDimNSE}, we define the vector $\bm{q}=(\nu,u,v,w,p)$ and write
\begin{align}
\begin{split}
    \frac{\partial\bm{q}}{\partial t}
    &
    +[A_{x}(\bm{q})+B_{x}(\bm{q})]\frac{\partial\bm{q}}{\partial x}
    +[A_{y}(\bm{q})+B_{y}(\bm{q})]\frac{\partial\bm{q}}{\partial y}
    +[A_{z}(\bm{q})+B_{z}(\bm{q})]\frac{\partial\bm{q}}{\partial z}\\
    &+B_{xx}(\bm{q})\frac{\partial^2\bm{q}}{\partial x^2}
    +B_{yy}(\bm{q})\frac{\partial^2\bm{q}}{\partial y^2}
    +B_{zz}(\bm{q})\frac{\partial^2\bm{q}}{\partial z^2}
    +B_{xy}(\bm{q})\frac{\partial^2\bm{q}}{\partial x\partial y}
    +B_{xz}(\bm{q})\frac{\partial^2\bm{q}}{\partial x\partial z}
    +B_{yz}(\bm{q})\frac{\partial^2\bm{q}}{\partial y\partial z}=0.
\end{split}
\label{02:eq:opForm}
\end{align}
where $B$ denotes viscous terms while $A$ denotes inviscid terms. Next we decompose the flow into a time-invariant equilibrium solution, $\bar{\bm{q}}$, and a time-varying disturbance variable, $\bm{q}'$, such that $\bm{q}=\bar{\bm{q}}+\bm{q}'$, which we use to obtain
\begin{align}
\begin{split}
    \frac{\partial\bm{q}'}{\partial t}
    &
    +[A_{x}(\bar{\bm{q}})+B_{x}(\bar{\bm{q}})]\frac{\partial\bm{q}'}{\partial x}
    +[A_{y}(\bar{\bm{q}})+B_{y}(\bar{\bm{q}})]\frac{\partial\bm{q}'}{\partial y}
    +[A_{z}(\bar{\bm{q}})+B_{z}(\bar{\bm{q}})]\frac{\partial\bm{q}'}{\partial z}
    +C(\bar{\bm{q}})\bm{q}'\\
    &
    +B_{xx}(\bar{\bm{q}})\frac{\partial^2\bm{q}'}{\partial x^2}
    +B_{yy}(\bar{\bm{q}})\frac{\partial^2\bm{q}'}{\partial y^2}
    +B_{zz}(\bar{\bm{q}})\frac{\partial^2\bm{q}'}{\partial z^2}
    +B_{xy}(\bar{\bm{q}})\frac{\partial^2\bm{q}'}{\partial x\partial y}
    +B_{xz}(\bar{\bm{q}})\frac{\partial^2\bm{q}'}{\partial x\partial z}
    +B_{yz}(\bar{\bm{q}})\frac{\partial^2\bm{q}'}{\partial y\partial z}=\bm{F}(\bm{q}'),
\end{split}
\label{02:eq:opFormNLin}
\end{align}
where we have defined $C(\bar{\bm{q}})$ such that
\begin{align}
\begin{split}
    C(\bar{\bm{q}})\bm{q}'
    &
    =[A_x(\bm{q}')+B_x(\bm{q}')]\frac{\partial\bar{\bm{q}}}{\partial x}
    +[A_y(\bm{q}')+B_y(\bm{q}')]\frac{\partial\bar{\bm{q}}}{\partial y}
    +[A_z(\bm{q}')+B_z(\bm{q}')]\frac{\partial\bar{\bm{q}}}{\partial z}\\
    &
    +B_{xx}(\bm{q}')\frac{\partial^2\bar{\bm{q}}}{\partial x^2}
    +B_{yy}(\bm{q}')\frac{\partial^2\bar{\bm{q}}}{\partial y^2}
    +B_{zz}(\bm{q}')\frac{\partial^2\bar{\bm{q}}}{\partial z^2}
    +B_{xy}(\bm{q}')\frac{\partial^2\bar{\bm{q}}}{\partial x\partial y}
    +B_{xz}(\bm{q}')\frac{\partial^2\bar{\bm{q}}}{\partial x\partial z}
    +B_{yz}(\bm{q}')\frac{\partial^2\bar{\bm{q}}}{\partial y\partial z},
\end{split}
\end{align}
and the nonlinear term
\begin{align}
\begin{split}
    \bm{F}(\bm{q}')
    &
    =
    -[A_x(\bm{q}')+B_x(\bm{q}')]\frac{\partial\bm{q}'}{\partial x}
    -[A_y(\bm{q}')+B_y(\bm{q}')]\frac{\partial\bm{q}'}{\partial y}
    -[A_z(\bm{q}')+B_z(\bm{q}')]\frac{\partial\bm{q}'}{\partial z}\\
    &
    -B_{xx}(\bm{q}')\frac{\partial^2\bm{q}'}{\partial x^2}
    -B_{yy}(\bm{q}')\frac{\partial^2\bm{q}'}{\partial y^2}
    -B_{yy}(\bm{q}')\frac{\partial^2\bm{q}'}{\partial z^2}
    -B_{xy}(\bm{q}')\frac{\partial^2\bm{q}'}{\partial x\partial y}
    -B_{xz}(\bm{q}')\frac{\partial^2\bm{q}'}{\partial x\partial z}
    -B_{yz}(\bm{q}')\frac{\partial^2\bm{q}'}{\partial y\partial z}.
\end{split}
\end{align}
We further define the linear operator
\begin{align}
\begin{split}
    \mathcal{L}(\bar{\bm{q}})
    =&
    -\frac{\partial }{\partial t}
    -[A_{y}(\bar{\bm{q}})+B_{y}(\bar{\bm{q}})]\frac{\partial\bm{q}'}{\partial y}
    -[A_{z}(\bar{\bm{q}})+B_{z}(\bar{\bm{q}})]\frac{\partial\bm{q}'}{\partial z}
    -C(\bar{\bm{q}})\bm{q}'\\
    &-B_{yy}(\bar{\bm{q}})\frac{\partial^2\bm{q}'}{\partial y^2}
    -B_{zz}(\bar{\bm{q}})\frac{\partial^2\bm{q}'}{\partial z^2}
    -B_{yz}(\bar{\bm{q}})\frac{\partial^2\bm{q}'}{\partial y\partial z}=0,
    \end{split}
\end{align}
yielding
\begin{equation}
    A_{x}(\bar{\bm{q}})\frac{\partial\bm{q}'}{\partial x}
    = \mathcal{L}(\bar{\bm{q}})\bm{q}'
    + \bm{F}(\bm{q}')
    - B_{x}(\bar{\bm{q}})\frac{\partial\bm{q}'}{\partial x}
    - B_{xx}(\bar{\bm{q}})\frac{\partial^2\bm{q}'}{\partial x^2}
    - B_{xy}(\bar{\bm{q}})\frac{\partial^2\bm{q}'}{\partial x\partial y}
    - B_{xz}(\bar{\bm{q}})\frac{\partial^2\bm{q}'}{\partial x\partial z}.
    \label{eq:operatorForm}
\end{equation}

The operators for the first derivatives (without viscous terms), $A_{x}$, $A_{y}$, and $A_{z}$, are given by
\[
A_{x}(\bm{q})=\left[\begin{array}{ccccc}
u & -\nu & 0 & 0 & 0\\
0 & u & 0 & 0 & \nu\\
0 & 0 & u & 0 & 0\\
0 & 0 & 0 & u & 0\\
0 & \gamma p & 0 & 0 & u
\end{array}\right],\quad
A_{y}(\bm{q})=\left[\begin{array}{ccccc}
v & 0 & -\nu & 0 & 0\\
0 & v & 0 & 0 & 0\\
0 & 0 & v & 0 & \nu\\
0 & 0 & 0 & v & 0\\
0 & 0 & \gamma p & 0 & v
\end{array}\right],\quad
A_{z}(\bm{q})=\left[\begin{array}{ccccc}
w & 0 & 0 & -\nu & 0\\
0 & w & 0 & 0 & 0\\
0 & 0 & w & 0 & 0\\
0 & 0 & 0 & w & \nu\\
0 & 0 & 0 & \gamma p & w
\end{array}\right],
\]
while the operators for the first derivatives (with viscous terms) $B_{x}$, $B_{y}$, and $B_{z}$, are given by
\[
B_{x}(\bm{q})=-\frac{2\gamma}{Pr Re}
\left[\begin{array}{ccccc}
0 & 0 & 0 & 0 & 0\\
0 & 0 & 0 & 0 & 0\\
0 & 0 & 0 & 0 & 0\\
0 & 0 & 0 & 0 & 0\\
\frac{\partial p}{\partial x} & 0 & 0 & 0 & \frac{\partial\nu}{\partial x}
\end{array}\right],\quad
B_{y}(\bm{q})=-\frac{2\gamma}{Pr Re}
\left[\begin{array}{ccccc}
0 & 0 & 0 & 0 & 0\\
0 & 0 & 0 & 0 & 0\\
0 & 0 & 0 & 0 & 0\\
0 & 0 & 0 & 0 & 0\\
\frac{\partial p}{\partial y} & 0 & 0 & 0 & \frac{\partial\nu}{\partial y}
\end{array}\right],
\]
and
\[
B_{z}(\bm{q})=-\frac{2\gamma}{Pr Re}
\left[\begin{array}{ccccc}
0 & 0 & 0 & 0 & 0\\
0 & 0 & 0 & 0 & 0\\
0 & 0 & 0 & 0 & 0\\
0 & 0 & 0 & 0 & 0\\
\frac{\partial p}{\partial z} & 0 & 0 & 0 & \frac{\partial\nu}{\partial z}
\end{array}\right],
\]
For the second derivatives $B_{xx}$ and $B_{yy}$ we have
\[
B_{xx}(\bm{q})
=\left[\begin{array}{ccccc}
0 & 0 & 0 & 0 & 0\\
0 & -\frac{4}{3}\frac{\nu}{Re} & 0 & 0 & 0\\
0 & 0 & -\frac{\nu}{Re} & 0 & 0\\
0 & 0 & 0 & -\frac{\nu}{Re} & 0\\
-\frac{\gamma p}{RePr} & 0 & 0 & 0 & -\frac{\gamma\nu}{RePr}
\end{array}\right],\quad
B_{yy}(\bm{q})
=\left[\begin{array}{ccccc}
0 & 0 & 0 & 0 & 0\\
0 & -\frac{\nu}{Re} & 0 & 0 & 0\\
0 & 0 & -\frac{4}{3}\frac{\nu}{Re} & 0 & 0\\
0 & 0 & 0 & -\frac{\nu}{Re} & 0\\
-\frac{\gamma p}{RePr} & 0 & 0 & 0 & -\frac{\gamma\nu}{RePr}
\end{array}\right],
\]
while for $B_{zz}$ and $B_{xy}$ we have
\[
B_{zz}(\bm{q})
=\left[\begin{array}{ccccc}
0 & 0 & 0 & 0 & 0\\
0 & -\frac{\nu}{Re} & 0 & 0 & 0\\
0 & 0 & -\frac{\nu}{Re} & 0 & 0\\
0 & 0 & 0 & -\frac{4}{3}\frac{\nu}{Re} & 0\\
-\frac{\gamma p}{RePr} & 0 & 0 & 0 & -\frac{\gamma\nu}{RePr}
\end{array}\right],\quad
B_{xy}(\bm{q})
=\left[\begin{array}{ccccc}
0 & 0 & 0 & 0 & 0\\
0 & 0 & -\frac{\nu}{3Re} & 0 & 0\\
0 & -\frac{\nu}{3Re} & 0 & 0 & 0\\
0 & 0 & 0 & 0 & 0\\
0 & 0 & 0 & 0 & 0
\end{array}\right],
\]
and for $B_{xz}$ and $B_{yz}$ we have
\[
B_{xz}(\bm{q})
=\left[\begin{array}{ccccc}
0 & 0 & 0 & 0 & 0\\
0 & 0 & 0 & -\frac{\nu}{3Re} & 0\\
0 & 0 & 0 & 0 & 0\\
0 & -\frac{\nu}{3Re} & 0 & 0 & 0\\
0 & 0 & 0 & 0 & 0
\end{array}\right],\quad
B_{yz}(\bm{q})
=\left[\begin{array}{ccccc}
0 & 0 & 0 & 0 & 0\\
0 & 0 & 0 & 0 & 0\\
0 & 0 & 0 & -\frac{\nu}{3Re} & 0\\
0 & 0 & -\frac{\nu}{3Re} & 0 & 0\\
0 & 0 & 0 & 0 & 0
\end{array}\right].
\]
Finally, the operator $C$ is given by
\[
C(\bm{q})
=\left[\begin{array}{ccccc}
-\nabla\cdot\bm{u} &
\frac{\partial\nu}{\partial x} &
\frac{\partial\nu}{\partial y} &
\frac{\partial\nu}{\partial z} &
0\\
\frac{\partial p}{\partial x}-\frac{1}{Re}\nabla^{2}u-\frac{1}{3Re}[\partial_{xx}u+\partial_{xy}v+\partial_{xz}w] &
\frac{\partial u}{\partial x} &
\frac{\partial u}{\partial y} &
\frac{\partial u}{\partial z} &
0\\
\frac{\partial p}{\partial y}-\frac{1}{Re}\nabla^{2}v-\frac{1}{3Re}[\partial_{xy}u+\partial_{yy}v+\partial_{yz}w] &
\frac{\partial v}{\partial x} &
\frac{\partial v}{\partial y} &
\frac{\partial v}{\partial z} &
0\\
\frac{\partial p}{\partial z}-\frac{1}{Re}\nabla^{2}w-\frac{1}{3Re}[\partial_{xz}u+\partial_{yz}v+\partial_{zz}w] &
\frac{\partial w}{\partial x} &
\frac{\partial w}{\partial y} &
\frac{\partial w}{\partial z} &
0\\
-\frac{\gamma}{RePr}\nabla^{2}p &
\frac{\partial p}{\partial x} &
\frac{\partial p}{\partial y} &
\frac{\partial p}{\partial z} &
\gamma\nabla\cdot\bm{u}-\frac{\gamma}{RePr}\nabla^{2}\nu
\end{array}\right],
\]
while the nonlinear term is a vector comprising the following components
\begin{align*}
    F_1(\bm{q}) &
    =u\frac{\partial\nu}{\partial x}
    +v\frac{\partial\nu}{\partial y}
    +w\frac{\partial\nu}{\partial z}
    -\nu\frac{\partial u}{\partial x}
    -\nu\frac{\partial v}{\partial y}
    -\nu\frac{\partial w}{\partial z},\\
    F_2(\bm{q}) &
    = 
    \nu\frac{\partial p}{\partial x}
    -u\frac{\partial u}{\partial x}
    -v\frac{\partial u}{\partial y}
    -w\frac{\partial u}{\partial z}
    +\frac{1}{Re}\nu\nabla^2 u
    +\frac{1}{3Re}\Big(\frac{\partial^2 u}{\partial x^2}+\frac{\partial^2 v}{\partial x\partial y}
    +\frac{\partial^2 w}{\partial x\partial z}\Big),\\
    F_3(\bm{q}) &
    = 
    \nu\frac{\partial p}{\partial y}
    -u\frac{\partial v}{\partial x}
    -v\frac{\partial v}{\partial y}
    -w\frac{\partial v}{\partial z}
    +\frac{1}{Re}\nu\nabla^2 u
    +\frac{1}{3Re}\Big(\frac{\partial^2 u}{\partial x\partial y}+\frac{\partial^2 v}{\partial y^2}
    +\frac{\partial^2 w}{\partial y\partial z}\Big),\\
    F_4(\bm{q}) &
    = 
    \nu\frac{\partial p}{\partial z}
    -u\frac{\partial w}{\partial x}
    -v\frac{\partial w}{\partial y}
    -w\frac{\partial w}{\partial z}
    +\frac{1}{Re}\nu\nabla^2 u
    +\frac{1}{3Re}\Big(\frac{\partial^2 u}{\partial x\partial z}+\frac{\partial^2 v}{\partial y\partial z}
    +\frac{\partial^2 w}{\partial z^2}\Big),\\
    F_5(\bm{q}) &
    =\frac{\gamma}{Re Pr}[\nu\nabla^2 p+2\nabla p\cdot\nabla\nu+p\nabla^2 \nu]
    -\bm{u}\cdot\nabla p
    -\gamma p \nabla\cdot{\bm{u}}.
\end{align*}

\section{Jacobian of the NOWNS equations}\label{app:nownsJac}

Here we derive the Jacobian of the fully-discrete system of NOWNS equations~\eqref{02:eq:nlinSysFD} for 2D disturbances. However, this analysis readily extends to 3D disturbances as well. We first introduce the NOWNS residual
\begin{equation}
    \hat{\bm{r}}^{\ddagger(k+1)}_{m}
    =\sum_{l=0}^{s-1} c^{(l)}A^{\ddagger}\hat{\bm{\phi}}_{m}^{\ddagger(k+1-l)}
    -
    \hat{L}_{m}^\ddagger\hat{\bm{\phi}}_{m}^{\ddagger(k+1)}
    -\hat{\bm{F}}_{m}^{\ddagger(k+1)}
    -\hat{\bm{f}}_{m}^{\ddagger(k+1)}
    =0,\quad m=0,\dots,2M,
\end{equation}
and take it's derivative with respect to $\hat{\bm{\phi}}_{p}^{\ddagger(k+1)}$ to obtain
\begin{equation}
    \frac{\partial\hat{\bm{r}}^{\ddagger(k+1)}_{m}}{\partial\hat{\bm{\phi}}_{p}^{\ddagger(k+1)}}
    =
    (c^{(0)}A^{\ddagger}-\hat{L}_{m}^\ddagger)
    \delta_{mp}
    -\frac{\partial \hat{\bm{F}}_{m}^{\ddagger(k+1)}}
    {\partial \hat{\bm{\phi}}_{p}^{\ddagger(k+1)}},\quad
    m,p=0,\dots,2M,
\end{equation}
Using the definition DFT, it can be shown (for 2D disturbances) that
\begin{equation}
    \frac{\partial \hat{\bm{F}}^{\ddagger(k+1)}_{m}}{\partial \hat{\bm{\phi}}^{\ddagger(k+1)}_{p}}
    =\frac{1}{2M+1}\sum_{l=0}^{2M}
    \frac{\partial\tilde{\bm{F}}^{\ddagger(k+1)}}{\partial\bm{\phi}^{\ddagger(k+1)}}\Big|_{\bm{\phi}=\bm{\phi}_{l}}
    e^{i2\pi l\frac{p-m}{2M+1}},
\end{equation}
which allows us to write
\begin{equation}
    \hat{J}_{p-m}^{\ddagger}=
    -\frac{\partial \hat{\bm{F}}_{m}^{\ddagger(k+1)}}{\partial \hat{\bm{\phi}}_{p}^{\ddagger(k+1)}}
    =
    -
    \begin{bmatrix}
        0 & 0 & 0\\
        P_{1,m}
        \begin{bmatrix}
        \tilde{A}_{\pm\pm}^{-1}\hat{J}_{\pm\pm,p-m}\\
        \hat{J}_{0\pm,p-m}
        \end{bmatrix}
        &
        P_{1,m}
        \begin{bmatrix}
        \tilde{A}_{\pm\pm}^{-1}\hat{J}_{\pm0,p-m}\\
        \hat{J}_{00,p-m}
        \end{bmatrix}
        & 0
        \\
        \hat{J}_{0\pm,p-m} & 
        \hat{J}_{00,p-m} & 0
    \end{bmatrix},\quad
    m,p=0,\dots,2M.
\end{equation}
We drop the $(k+1)$ superscript for simplicity, and for $M=1$ we can write
\begin{equation}
\begin{bmatrix}
        (c^{(0)}A^{\ddagger}-\hat{L}_{0}^\ddagger)
        + \hat{J}^{\ddagger}_{0} &
        \overline{\hat{J}^{\ddagger}_{1}} &
        \hat{J}^{\ddagger}_{1} \\
        \hat{J}^{\ddagger}_{1} &
        (c^{(0)}A^{\ddagger}
        -\hat{L}^{\ddagger}_{1})
        + \hat{J}^{\ddagger}_{0} &
        \overline{\hat{J}^{\ddagger}_{1}} \\
        \overline{\hat{J}^{\ddagger}_{1}} &
        \hat{J}^{\ddagger}_{1} &
        (c^{(0)}A^{\ddagger}
        -\hat{L}_{-1}^\ddagger) + \hat{J}^{\ddagger}_{0}
    \end{bmatrix}
    \begin{bmatrix}
        \Delta\hat{\bm{\phi}}_0^\ddagger \\
        \Delta\hat{\bm{\phi}}_1^\ddagger \\
        \overline{\Delta\hat{\bm{\phi}}_1}^\ddagger
    \end{bmatrix}
    =
    \begin{bmatrix}
        \hat{r}_0^\ddagger \\
        \hat{r}_1^\ddagger \\
        \overline{\hat{r}_1}^\ddagger
    \end{bmatrix}.
    \label{eq:newton}
\end{equation}
where we have used $\hat{J}^{\ddagger}_{m}=\hat{J}^{\ddagger}_{m+2M+1}$ and $\hat{J}^{\ddagger}_{m}=\overline{\hat{J}^{\ddagger}_{-m}}$ for $m=0,\dots,M$.


\section{Streamwise diffusion terms}\label{sec:streamwiseDiffStudy}

\begin{figure}
     \centering
     \begin{subfigure}[b]{0.45\textwidth}
         \centering
         \includegraphics[width=\textwidth]{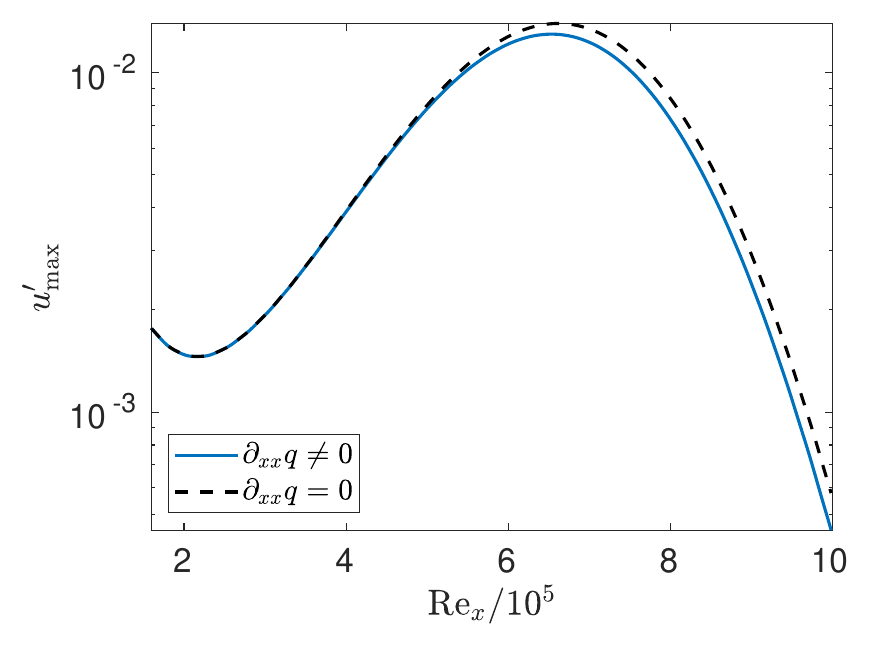}
         \caption{Linear calculation}
         \label{fig:2D-diffusion-lin}
     \end{subfigure}
     \hfill
     \begin{subfigure}[b]{0.45\textwidth}
         \centering
         \includegraphics[width=\textwidth]{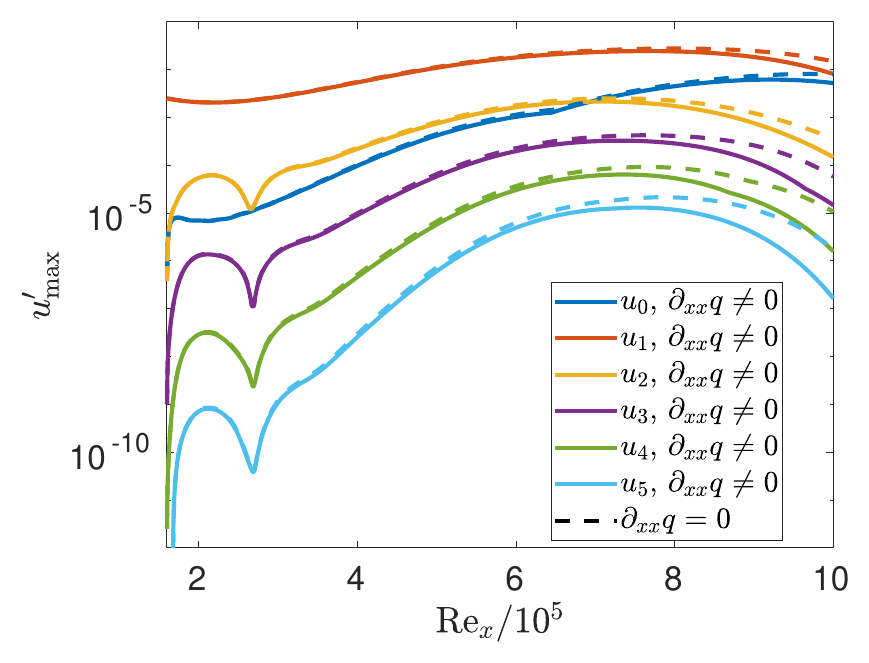}
         \caption{Nonlinear calculation}
         \label{fig:2D-diffusion-nlin}
     \end{subfigure}     
    \caption{$u$-velocity amplitudes with and without streamwise diffusion terms for the 2D evolution of the TS wave.}
    \label{fig:2D-diffusion}
\end{figure}

Here we demonstrate for the 2D validation case discussed in section~\ref{sec:results-2D} that although streamwise diffusion effects have a minimal impact on the linear calculation (figure~\ref{fig:2D-diffusion-lin}), the impact is more pronounced in the nonlinear case (figure~\ref{fig:2D-diffusion-nlin}). We see that prior to the second neutral stability point, the calculations with and without the streamwise diffusion terms yield similar amplitudes, but that downstream of this point, the amplitudes of the calculation including the streamwise viscous terms has much lower amplitudes (particularly for the higher harmonics in the nonlinear case).

\section{Inclusion of the streamwise pressure gradient}\label{sec:dp}

\begin{figure}
     \centering
     \begin{subfigure}[b]{0.45\textwidth}
         \centering
         \includegraphics[width=\textwidth]{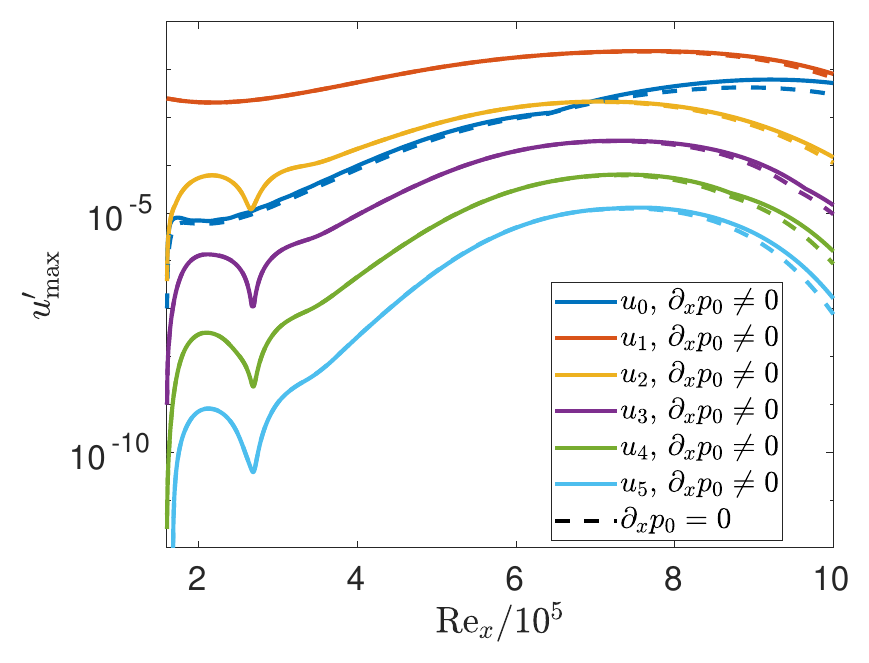}
         \caption{$u$-velocity amplitudes}
         \label{fig:2D-pressure-amplitudes}
     \end{subfigure}
     \hfill
     \begin{subfigure}[b]{0.45\textwidth}
         \centering
         \includegraphics[width=\textwidth]{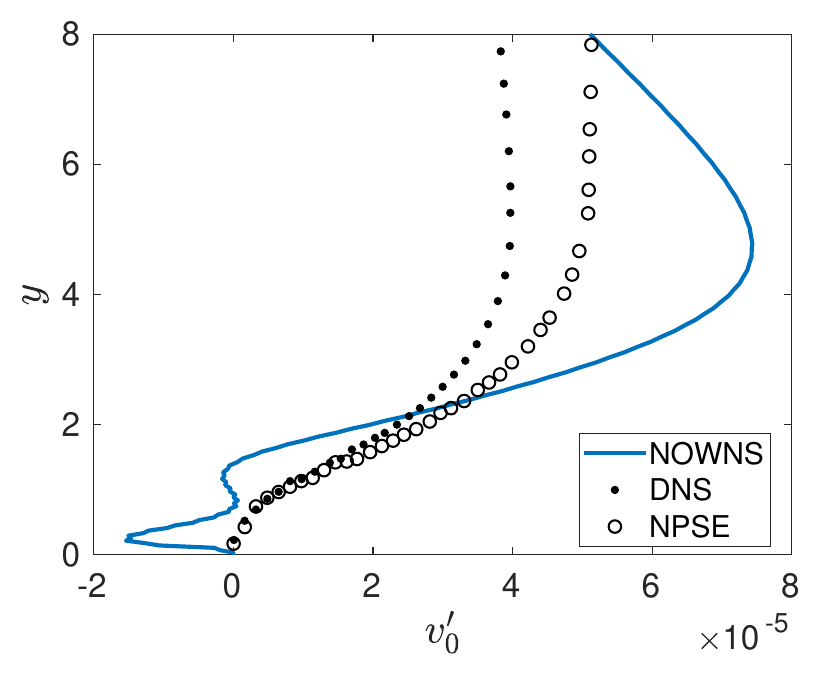}
         \caption{$v$-velocity profile}
         \label{fig:2D-pressure-profile-v0}
     \end{subfigure}     
    \caption{2D validation case with and without streamwise pressure gradient for zero-frequency modes.}
    \label{fig:2D-pressure}
\end{figure}

Unlike the NPSE approach, the NOWNS approach has no minimum step size requirement for the march to be numerically stable, but it is still necessary to neglect the streamwise pressure gradient for the zero-frequency modes. If the streamwise pressure gradient for these modes is retained, the NOWNS march remains stable but becomes inaccurate, as demonstrated in figure~\ref{fig:2D-pressure}. We plot the $u$-velocity amplitudes with and without $\partial_x p_0$ in figure~\ref{fig:2D-pressure-amplitudes}, where we see that we have reasonable agreement for $m\neq0$, but disagreement for $m=0$. In figure~\ref{fig:2D-pressure-profile-v0} we plot the profile of the $v_0$, and we notice that the profile predicted by NOWNS when $\partial_x p_0$ is included is substantially different from the profiles predicted by NPSE and DNS.

If we neglect $\partial_x p_0$, but include the streamwise diffusion terms $\partial_{xx} q_0$, then we must project the MFD and we find that we have good agreement between the DNS and NOWNS calculations. Therefore, we can conclude that the recursion parameters we are using for the zero-frequency modes are valid. However, as discussed above, the calculation remains stable but becomes inaccurate when we include $\partial_x p_0$. Although our march remains stable, there including $\partial_x p_0$ leads to inaccuracies in the march for unknown reasons.

\begin{figure}
     \centering
     \begin{subfigure}[b]{0.45\textwidth}
         \centering
         \includegraphics[width=\textwidth]{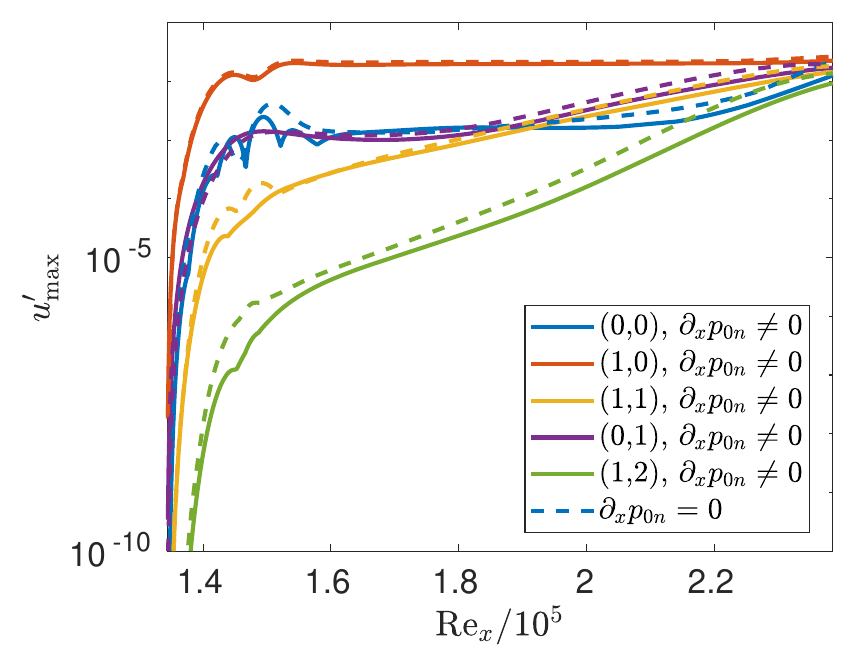}
         \caption{$u$-velocity amplitudes}
         \label{fig:Ktype_Pressure}
     \end{subfigure}
     \hfill
     \begin{subfigure}[b]{0.45\textwidth}
         \centering
         \includegraphics[width=\textwidth]{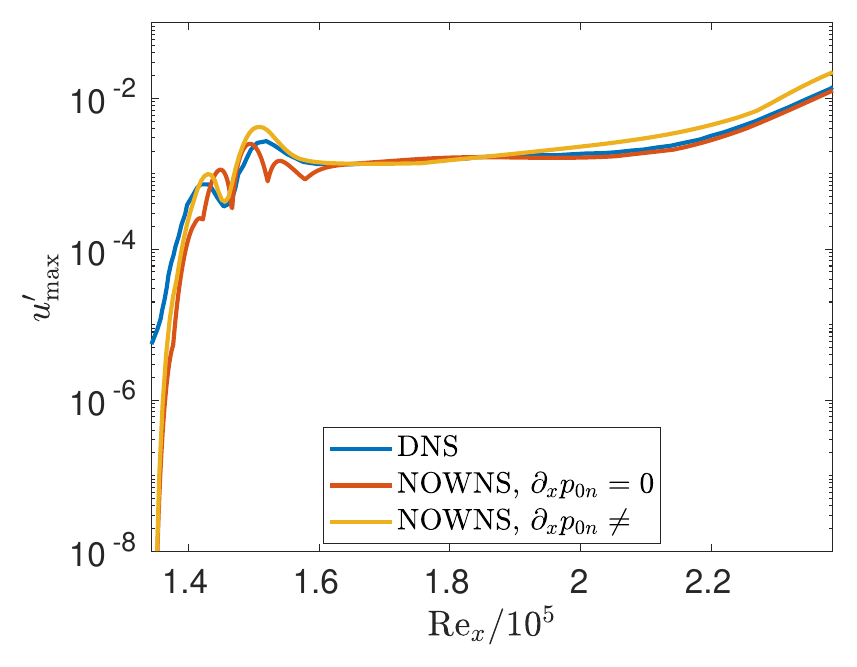}
         \caption{$u$-velocity amplitudes for MFD}
         \label{fig:Ktype_MFD}
     \end{subfigure}     
    \caption{K-type transition case with and without streamwise pressure gradient for zero-frequency modes.}
    \label{fig:Ktype_PressureCombined}
\end{figure}

In figure~\ref{fig:Ktype_Pressure}, we plot the $u$-velocity amplitudes computed by NOWNS with and without the pressure gradient for the zero-frequency modes. Mode (1,0) is tracked reasonably accurately, but the other modes are not. In particularly, the (1,2), (1,1), and (0,1) modes and the MFD have higher amplitudes than they should. In figure~\ref{fig:Ktype_MFD}, we see that we have better qualitative agreement in the early stages of the march between the NOWNNS and DNS calculations when the streamwise pressure gradient terms are include for the zero-frequency modes. However, we have worse quantitative agreement in the later stages of the march, which in turn causes the larger amplitudes of the (1,2), (1,1) and (0,1) modes observed in figure~\ref{fig:Ktype_Pressure}.

\section{Comparison of nonlinear solution procedures}\label{sec:nonlinSolveStudy}

\begin{figure}
     \centering
     \begin{subfigure}[b]{0.45\textwidth}
         \centering
        \includegraphics[width=\textwidth]{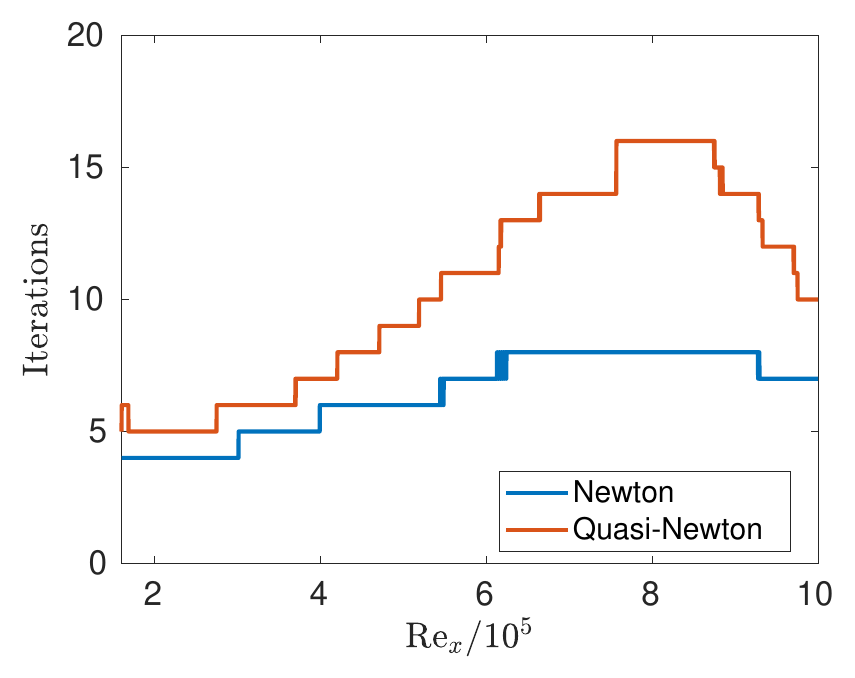}
        \caption{Comparison of Newton's method and the quasi-Newton method for the 2D validation case}
        \label{fig:2D-convergence}
     \end{subfigure}
     \hfill
     \begin{subfigure}[b]{0.45\textwidth}
         \centering
         \includegraphics[width=\textwidth]{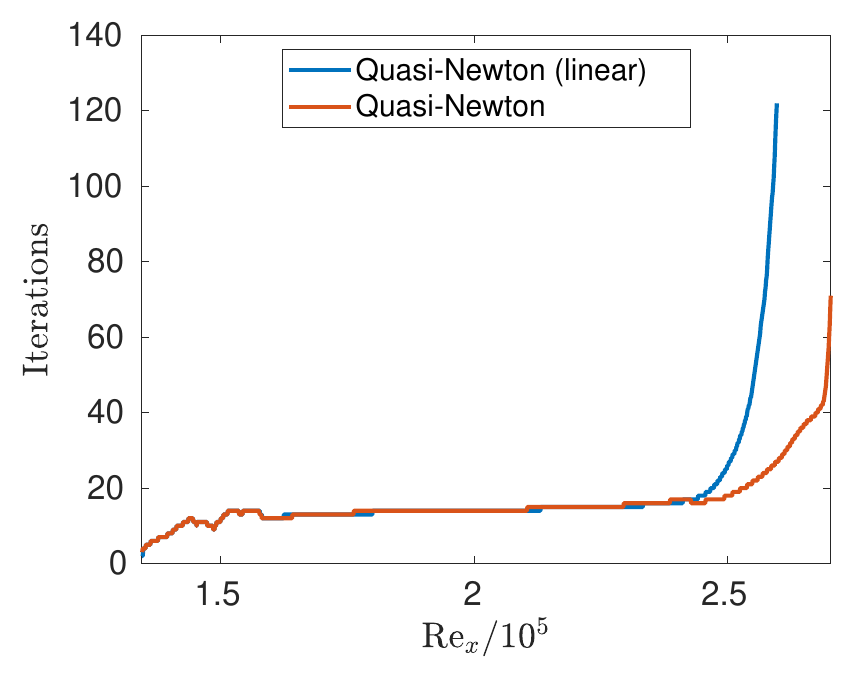}
         \caption{Comparison of the linear and nonlinear quasi-Newton methods for K-type transition}
         \label{fig:Ktype_Convergence}
     \end{subfigure}     
    \caption{Comparison of solution procedures for the nonlinear system of equations}
    \label{fig:convergence}
\end{figure}

We have discussed three procedures for solving the nonlinear system of equations: (i) Newton's method, (ii) a quasi-Newton method that includes part of the nonlinear Jacobian, and (iii) a quasi-Newton method that excludes the Jacobian of the nonlinear term. We plot the iterations to converge as a function of streamwise station in figure~\ref{fig:convergence}. For the 2D validation case, we compare the iterations to convergence for the quasi-Newton method and Newton's method in figure~\ref{fig:2D-convergence}, which shows that Newton's method converges in fewer iterations. In figure~\ref{fig:Ktype_Convergence}, we make the same comparison for linear and nonlinear quasi-Newton methods for K-type transition. We see that farther downstream, where the nonlinearity is stronger, the nonlinear quasi-Newton method converges in fewer iterations than the linear version. However, for sufficiently strong nonlinearities, it is necessary to use Newton's method.

\section{Linearization about the baseflow vs.\ the corrected mean flow}\label{sec:meanFlowStudy}

\begin{figure}
    \centering
    \includegraphics[width=0.6\columnwidth]{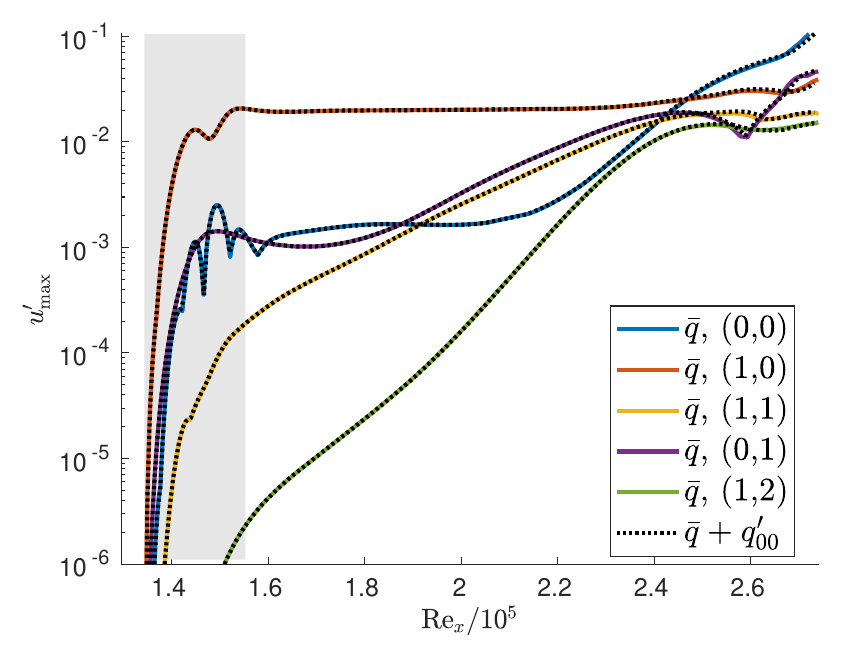}
    \caption{Amplitude of $u'$ v.s.\ streamwise coordinate, $\mathrm{Re}_x$, for fundamental breakdown. Compare linearizing NOWNS about $\bar{\bm{q}}$ vs.\ $\bar{\bm{q}}+\bm{q}'_{00}$.}
    \label{fig:ff_relin_amplitudes}
\end{figure}

Whereas in linear stability analysis, the mean flow is determined only by the baseflow, $\bar{\bm{q}}$, in the nonlinear case the disturbances interact to excite the MFD, $\bm{q}_{00}'$, which yields the corrected mean flow, $\bar{\bm{q}}+\bm{q}_{00}'$. In linear OWNS, we linearize the projection operators about the baseflow, while in nonlinear OWNS, we can choose to instead linearize about the corrected mean flow. For the K-type transition case of section~\ref{sec:kType}, we perform the NOWNS calculation again, but we instead linearize about the corrected mean flow. In figure~\ref{fig:ff_relin_amplitudes}, we compare the amplitudes computed by the two approaches to NOWNS and we see that in the early stages of the march, when the amplitudes are small, the two calculations are nearly identical, but that they begin to differ slightly as the disturbance amplitudes increase. The differences are small, but are most notable for the MFD, $\bm{q}_{00}'$, and the vortex mode, $q_{01}'$, after $\mathrm{Re}_x=2.5\times10^5$, which corresponds to an MFD amplitude of roughly 3\% of $U_{\infty}$. The two NOWNS marches continue to yield similar results until the end of the calculation at $\mathrm{Re}_x=2.74\times10^5$, where the MFD amplitude is 11\% of $U_{\infty}$.

Linearizing about the baseflow is more computationally efficient because the projection operators do not change between iterations. Moreover, the choice to linearize about the corrected mean flow instead of the baseflow does not appear to substantially affect the stability calculations, for MFD amplitudes less than roughly 10\% of $U_\infty$. Therefore, we recommend that NOWNS calculations be linearized about the baseflow rather than the corrected mean flow.

\section{Recursion parameter sets}\label{app:recursions}

The recursion parameter sets for non-zero frequency modes match those used in~\cite{Sleeman_2024_NOWNS-Aviation}. They depend on $\omega$ through the streamwise wave number $k=\omega/\bar{c}$, and when $k=0$, the recursion parameters associated with the vortical and propagating acoustic waves go to zero, leaving only the evanscent acoustic waves.

\section*{Acknowledgments}

This work has been supported by The Boeing Company through the Strategic Research and Development Relationship Agreement CT-BA-GTA-1.

\bibliography{ownslibrary}

\end{document}